%% file: paper.tex
\documentclass{aa}
\usepackage{textcomp}
\usepackage{graphicx}
\usepackage{amsmath}
\usepackage{amssymb}
\usepackage{bm}
\usepackage{upgreek}
\usepackage{IEEEtrantools}
\usepackage{multirow}
\usepackage[colorlinks=true,allcolors=blue,urlcolor=blue]{hyperref}
\usepackage{txfonts}
\usepackage[normalem]{ulem} 
\usepackage[dvipsnames]{xcolor}

\newcommand{\eqn}[1]{\text{Eq.~\ref{#1}}}
\newcommand{\sect}[1]{\text{Sect.~\ref{#1}}}
\newcommand{\fig}[1]{\text{Fig.~\ref{#1}}}
\newcommand{\tab}[1]{\text{Table~\ref{#1}}}

\newcommand{\mtd}{\textlangle3D\textrangle}

\newcommand{\scate}{\texttt{SCATE}}

\newcommand{\marcs}{\texttt{MARCS}}
\newcommand{\stagger}{\texttt{STAGGER}}
\newcommand{\cobold}{\texttt{CO$^{5}$BOLD}}
\newcommand{\atmo}{\texttt{ATMO}}
\newcommand{\handm}{HM74}

\newcommand{\kms}{\mathrm{km\,s^{-1}}}

\newcommand{\lggf}{\log{gf}}

\newcommand{\lgeps}[1]{\log{\epsilon_{\mathrm{#1}}}}

\newcommand{\dex}{\mathrm{dex}}

\newcommand{\nm}{\mathrm{nm}}
\newcommand{\K}{\mathrm{K}}

\newcommand{\eV}{\mathrm{eV}}
\newcommand{\vmic}{\xi_{\text{mic}}}

\newcommand{\iso}[2]{\mathrm{^{#1}#2}}
\newcommand{\ctwo}{C$_{2}$}
\newcommand{\comol}{$\iso{12}{C}\iso{16}{O}$}

\begin{document} 

\title{The solar carbon, nitrogen, and oxygen
abundances from a 3D LTE analysis of molecular lines\thanks{Table 2 is
available in electronic form
at the CDS via anonymous ftp to cdsarc.u-strasbg.fr (130.79.128.5)
or via \url{http://cdsarc.u-strasbg.fr/viz-bin/qcat?J/A+A/}}}
\author{A.~M.~Amarsi\inst{\ref{uu1}}
\and
N.~Grevesse\inst{\ref{liege1},\ref{liege2}}
\and
M.~Asplund
\and
R.~Collet\inst{\ref{aarhus}}}
\institute{\label{uu1}Theoretical Astrophysics, 
Department of Physics and Astronomy,
Uppsala University, Box 516, SE-751 20 Uppsala, Sweden\\
\email{anish.amarsi@physics.uu.se} 
\and 
\label{liege1}Centre Spatial de Li\`ege, Universit\'e de Li\'ege,
avenue Pr\'e Aily, 
B-4031 Angleur-Li\`ege, Belgium
\and
\label{liege2}Space sciences, Technologies and Astrophysics Research (STAR)
Institute, 
Universit\'e de Li\`ege, All\'ee du 6 ao\^ut, 17, B5C, 
B-4000 Li\`ege, Belgium
\and
\label{aarhus}Stellar Astrophysics Centre, 
Department of Physics and Astronomy, Aarhus University, 
Ny Munkegade 120, DK-8000 Aarhus C, Denmark}

\abstract{Carbon, nitrogen, and oxygen are the 
fourth, sixth, and third most abundant elements in the Sun. Their
abundances remain hotly debated due to the so-called solar modelling 
problem that has persisted for almost 20 years.
We revisit this issue by presenting a homogeneous analysis of 
$408$ molecular lines across $12$ diagnostic groups,
observed in the solar intensity spectrum.
Using a realistic 3D radiative-hydrodynamic model solar photosphere and LTE (local thermodynamic equilibrium) line formation,
we find $\lgeps{C}=8.47\pm0.02$, 
$\lgeps{N}=7.89\pm0.04$, and $\lgeps{O}=8.70\pm0.04$.
The stipulated uncertainties mainly reflect the sensitivity
of the results to the model atmosphere; this sensitivity is
correlated between the different diagnostic groups,
which all agree with the mean result to within $0.03\,\dex$.
For carbon and oxygen, 
the molecular results are in excellent agreement with 
our 3D non-LTE analyses of atomic lines.
For nitrogen, however, the molecular indicators give a 
$0.12\,\dex$ larger abundance than the atomic
indicators, and our best estimate of the 
solar nitrogen abundance is given by the mean: $7.83\,\dex$.
The solar oxygen abundance advocated here is close to our 
earlier determination of 
$8.69\,\dex$, and so the present results do not
significantly alleviate the solar modelling problem.}


\date{Received / Accepted}
\maketitle
\section{Introduction}
\label{introduction}

Carbon, nitrogen, and oxygen are 
among the most important elements in astrophysics today.
This is especially true in the context of our solar system,
where they are the fourth, sixth, and third most abundant elements
respectively \citep{2021arXiv210501661A}. Combined, they dominate the metal content of
the solar interior; as such their abundances are central to the debate on
the long-standing so-called solar modelling problem 
\citep[see e.g.][]{2005ApJ...621L..85B,
2008PhR...457..217B,2019FrASS...6...42B,2021LRSP...18....2C}. 
This describes an obstinate discrepancy between
helioseismic inferences versus first-principles modelling 
of the solar interior structure (as traced by the sound speed
as a function of depth, as well as quantities such as
the helium abundance of the solar envelope and 
the location of the bottom of the convection zone). 
The former approach is model-insensitive;
in contrast the latter approach depends the adopted description 
of the interior opacity, which in turn depends on the assumed
solar chemical composition. It has been suggested that
an approximately $15$\% larger opacity 
near the base of the convection zone, where oxygen transitions
are the largest contributors to the overall opacity together with iron
\citep{2015ApJS..220....2M},
could help to resolve this problem \citep{2009A&A...494..205C,
2009ApJ...705L.123S,2015Natur.517...56B}.

For the solar photosphere, absorption spectra of the molecules
\ctwo{}, CH, NH, OH, CN, and \comol{},
offer a promising way to infer the elemental abundances 
\citep{1978MNRAS.182..249L,1990A&A...232..225G,1991A&A...242..488G,
2004A&A...417..751A,2005A&A...431..693A}.
The molecules complement the usual atomic indicators of the
elemental abundances: namely, the high excitation
permitted lines of \ion{C}{I}, \ion{N}{I}, and \ion{O}{I}
that are typically more sensitive to departures from 
local thermodynamic equilibrium
(LTE; \citealt{1993A&A...275..269K,2005ARA&A..43..481A,2015A&A...583A..57S,2020A&A...642A..62A});
as well as the very weak low excitation forbidden lines
of \ion{C}{I} and \ion{O}{I} which form in LTE
but are heavily blended \citep{2008A&A...488.1031C,2019A&A...624A.111A}.
In contrast, the
molecular transitions are expected to be less sensitive to 
departures from LTE \citep{1975MNRAS.170..447H,1989ApJ...338.1033A}, 
and there are many unblended features in the solar spectrum to choose
from that can be measured with high precision.
However, the molecular concentrations are sensitive to the atmospheric structure and the presence of inhomogeneities, especially in high layers near the chromospheric temperature minimum,
which makes it essential to base the analysis on time-dependent,
three-dimensional (3D) radiative-hydrodynamical models of the photosphere
\citep{2004A&A...417..751A,2005A&A...431..693A}.
Historically, a disadvantage of using molecules
came from having to use rather uncertain energy levels and transition
probabilities; fortunately, 
driven in part by their importance in exoplanet modelling
\citep{2012MNRAS.425...21T,2016JMoSp.327...73T},
data of high accuracy are now available.

Results from molecules factor into 
the recommended carbon, nitrogen, and oxygen abundances
in the standard solar chemical composition of \citet{2009ARA&A..47..481A}.
However, there have been a number of developments since that work,
that make a new analysis worthwhile.
First, improved data have recently become available
for all of these molecules:
primarily transition probabilities
\citep{2013JQSRT.124...11B,
2014JChPh.141e4310B,2014ApJS..210...23B,
2015JChPh.143b6101B,2016JQSRT.168..142B,
2014A&A...571A..47M,2015ApJS..216...15L};
and also molecular
partition functions and equilibrium constants 
\citep{2016A&A...588A..96B}.
Secondly, a number of accurate, complementary
results have become available, and it is important
to verify that consistent results are obtained.
These include 3D non-LTE elemental abundances based on \ion{C}{I}
\citep{2019A&A...624A.111A,2021MNRAS.502.3780L},
\ion{N}{I} \citep{2020A&A...636A.120A}, 
and \ion{O}{I} \citep{2018A&A...616A..89A}.
Lastly,
the 3D model atmosphere has been updated,
as summarised in Table 1 of \citet{2021arXiv210501661A};
in particular, the newer model
was constructed with the composition of
\citep{2009ARA&A..47..481A}, rather than the more metal-poor
composition of \citep{2005ASPC..336...25A}.

Owing to these developments, we here present a selection and 
analysis of the best molecular diagnostics in the solar spectrum.
Preliminary results of the present study were already incorporated into
the new solar chemical composition of \citet{2021arXiv210501661A}.
We first present the selection of lines and their equivalent widths,
and review how the
3D LTE theoretical spectra were calculated (\sect{method}). 
We describe the analysis method and present the final
results (\sect{abundances}), and discuss them 
(\sect{discussion}), before presenting our conclusions
(\sect{conclusion}).

\section{Method}
\label{method}

\begin{figure*}
    \begin{center}
        \includegraphics[scale=0.31]{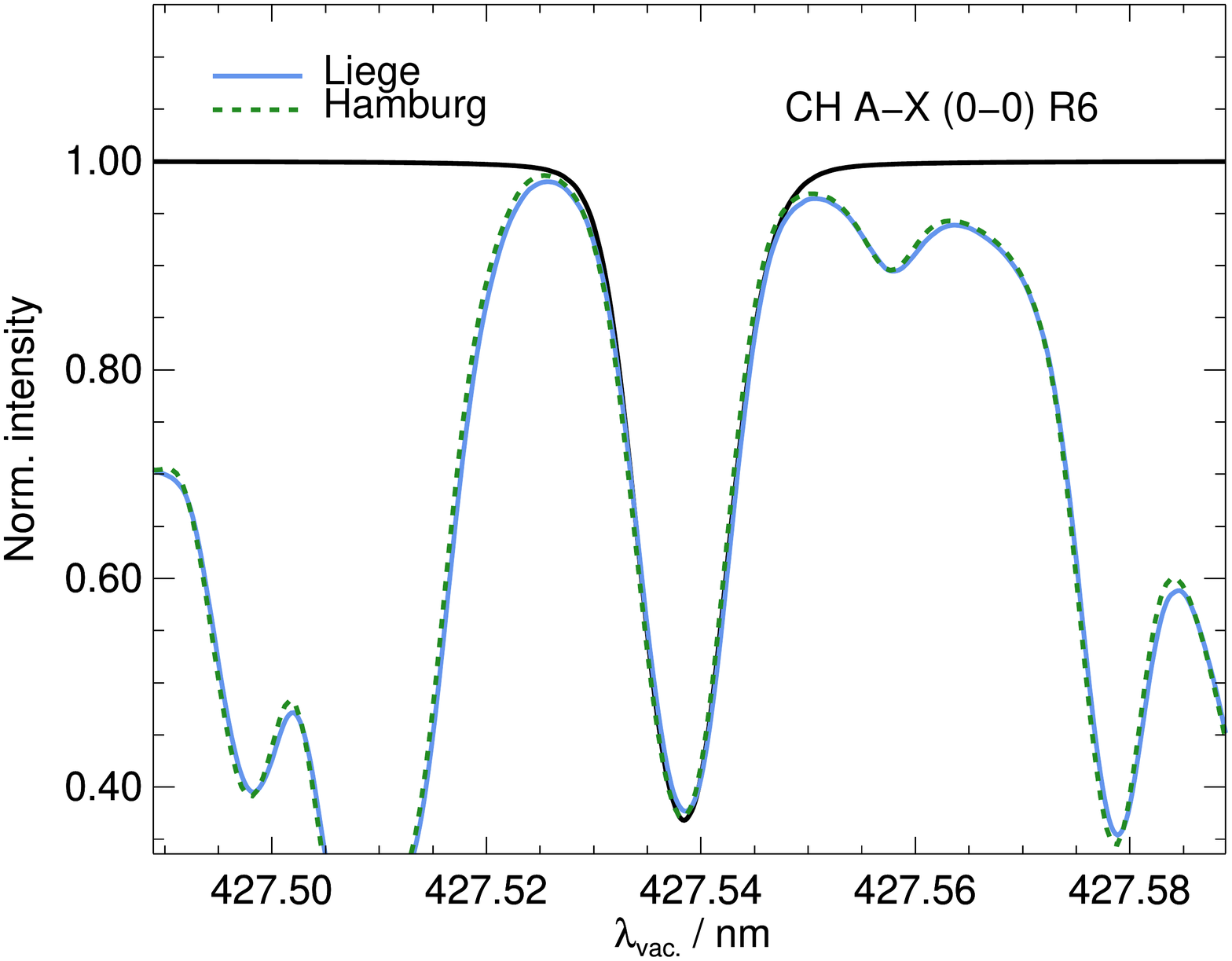}\includegraphics[scale=0.31]{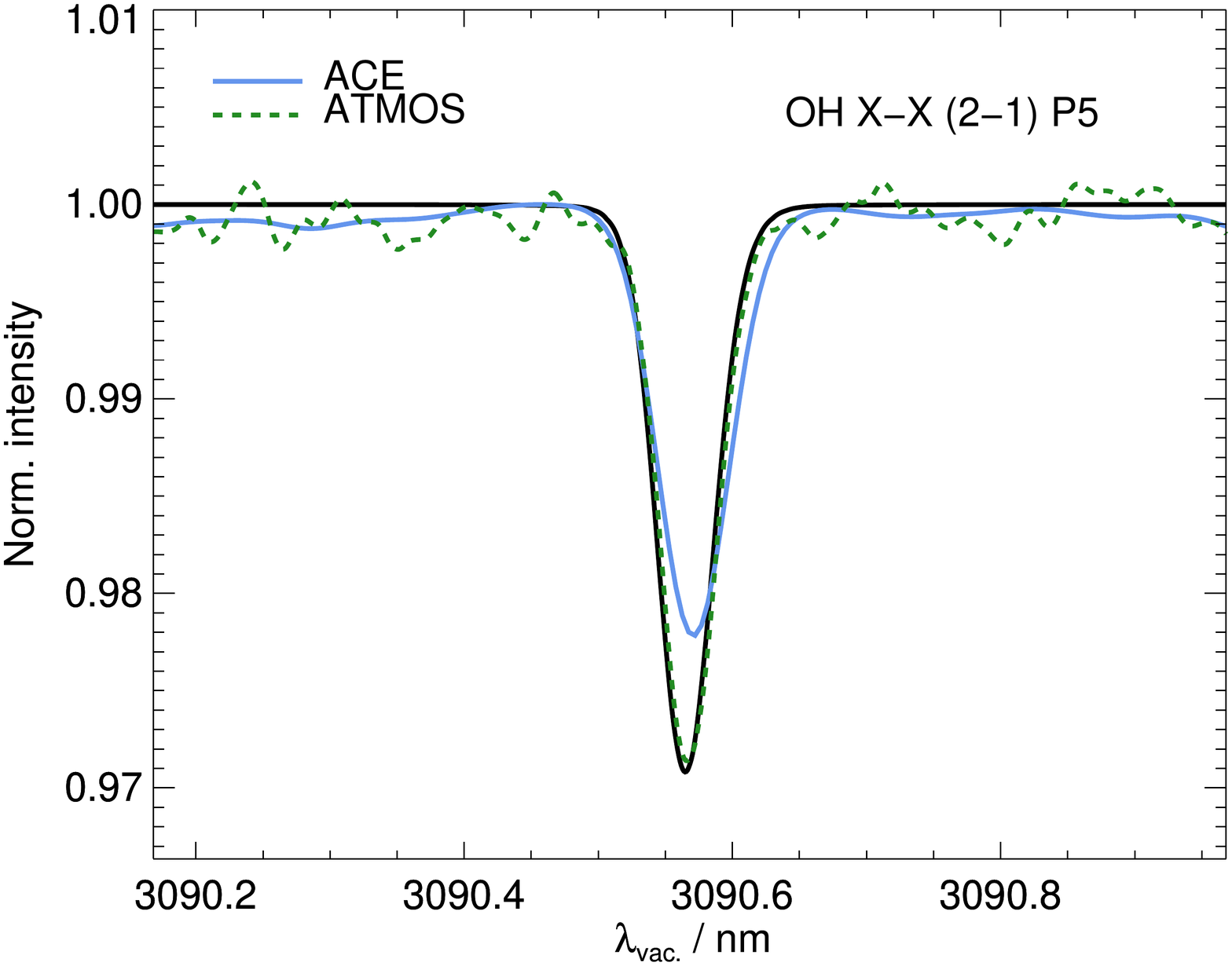}
        \caption{Illustrative example optical and infrared
        spectra from the 3D model
        compared to different solar atlases.
        The theoretical line strengths have not been fit to the data; rather,
        the input abundance
        is based on interpolating the theoretical equivalent-width
        to the mean one measured in both atlases.
        No instrumental broadening has been applied, and thus
        in the infrared
        the ACE profiles are seen to be broader and shallower
        than both the ATMOS and theoretical ones.}
        \label{fig:synthesis}
    \end{center}
\end{figure*}

\begin{figure}
    \begin{center}
        \includegraphics[scale=0.31]{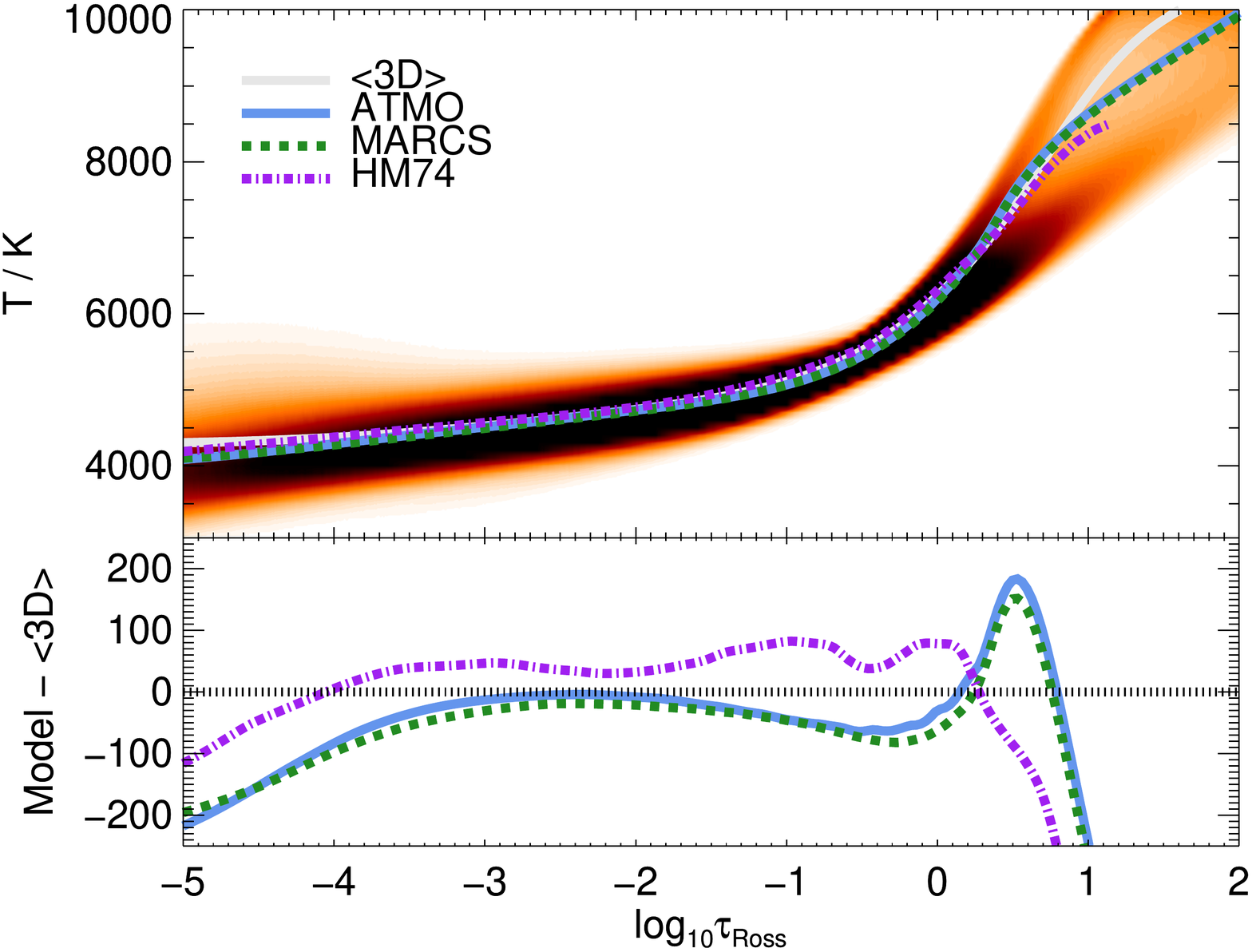}
        \caption{Distribution of temperature and
        Rosseland mean vertical optical depth 
        for the 3D radiative-hydrodynamical 
        model atmosphere, with darker colours indicating
        larger densities of grid points.
        The structures of the 1D model atmospheres used in this work
        are overplotted.
        Bottom panel shows the difference in temperature
        of the 1D models with respect to the horizontally 
        and temporally averaged \mtd{} model.}
        \label{fig:atmos}
    \end{center}
\end{figure}

\input{groups}

\input{linelist}

\subsection{Observational data and line selection}
\label{method_observations}

Our analysis is based on the solar disk-centre intensity spectrum. 
Different standard high-resolution solar atlases were used
for different lines, depending on their wavelengths.
For lines in the optical and near-infrared, 
the ground-based Li\`ege \citep{1973apds.book.....D,delbouille1981photometric,
1995ASPC...81...32D}
and Hamburg \citep{1984SoPh...90..205N,1999SoPh..184..421N}
atlases (see \citealt{2016A&A...590A.118D} for an
informative review and comparison of these atlases and others
in terms of e.g. spectral resolution,
telluric absorption and continuum placement) formed the observational basis.
For lines in the infrared,
the space-based ATMOS \citep{1996ApOpt..35.2747A}
and ACE \citep{2010JQSRT.111..521H} atlases were used.

In general, the equivalent widths of selected lines 
were determined separately in both atlases, by identifying
the continuum from local maxima within a few
tenths of a nanometre on either side of the central depression, 
and subsequently measuring the 
area under the continuum by hand using a planimeter.
The mean equivalent widths from the two atlases were adopted;
these are typically in 
excellent agreement, but can vary by up to around $5\%$ for problematic lines:
for instance, the \ctwo{} lines, which are typically weak and sit
in a crowded region of the optical spectrum, which makes the continuum
difficult to place and profile shapes difficult to trace.
The first overtone OH lines are an exception,
this analysis being based on
the equivalent widths of \citet{2004ApJ...615.1042M}, 
measured in the atlas of \citet{2003assi.book.....W}.
To give
some feeling for the spectral features analysed in this work,
we show a CH optical line and 
OH infrared line in \fig{fig:synthesis}.

In \tab{tab:groups} we summarise 
the different groups of lines considered, and the number of lines in each
group.  We present 
the entire list of line parameters
and their measured equivalent widths
in \tab{tab:linelist}.  These lines were chosen for being
free of blends and sufficiently strong to be measured reliably,
while remaining on or close to the linear part of the curve of growth, 
to enable the elemental abundances to be inferred with
high precision.
The wavelengths, energies, and transition probabilities were adopted from
\citet{2013JQSRT.124...11B,2014ApJS..210...23B,
2015JChPh.143b6101B,2016JQSRT.168..142B} for 
\ctwo{}, CN, NH, and OH respectively;
from \cite{2014A&A...571A..47M} for CH;
and from \citet{2015ApJS..216...15L} for \comol{}.

Many more lines were also considered, but not retained in the final
analysis.
Most notably, a further 463 lines of 14 other bands in the red system of CN
were studied and initially kept
\citep{2021arXiv210501661A}.
The analysis of these lines was based on equivalent widths that were measured
automatically over 20 years ago
for the purposes of line identification (A.~J.~Sauval, priv. comm.).
The mean results from these various bands were found to agree with
those from the $(0-0)$ band to within $0.05\,\dex$.
However, the dispersions of the results are two to three times larger
because of random uncertainties in the equivalent widths.
For this reason, these $463$ transitions were not retained in the final
analysis.
In addition, several lines in the
NH A-X,  OH A-X, and  CN B-X electronic systems were also considered.
Their lines fall in the near UV, around $340\,\nm$, 
$320\,\nm$, and $390\,\nm$,
respectively. There is a high density of spectral lines in these spectral
regions, making 
the precise measurement of equivalent widths
extremely difficult because of the
contamination by other species and because of the difficulty to find the real
continuum. For these reasons, 
these molecular transitions were ultimately not included.

\subsection{Model atmospheres}
\label{method_atmosphere}

In \fig{fig:atmos} we illustrate the temperature structure as a function of Rosseland optical depth of the model atmospheres used in this study.
The analysis was based on a 3D radiative-hydrodynamic model
of the solar photosphere computed with the \stagger{} code 
\citep{2011JPhCS.328a2003C,2013A&A...557A..26M}.
The mean effective temperature of the entire sequence
(as well as of the sample of snapshots; \sect{method_lineformation})
is $5773\,\K$, with a standard deviation of $16\,\K$, 
in excellent agreement with the nominal solar value of 
$5772\,\K$ \citep{2016AJ....152...41P}.
The model is the same as that
used already in the 3D non-LTE analyses of \ion{O}{I} lines
\citep{2018A&A...616A..89A}, where it is discussed in more detail;
as well as of \ion{C}{I} and \ion{N}{I} lines
\citep{2019A&A...624A.111A,2021MNRAS.502.3780L,2020A&A...636A.120A}.
It was also used in
the 3D non-LTE analyses of the sodium, magnesium, potassium, calcium,
and iron abundances, presented in
\citet{2021arXiv210501661A}.
It is an updated version of the model used in the solar analyses
of \citet{2009ARA&A..47..481A} as well as of 
\citet{2015A&A...573A..25S,2015A&A...573A..26S}
and \citet{2015A&A...573A..27G}, using the latest version of the 
\stagger{} code as described by \citet{2018MNRAS.475.3369C},
with an updated chemical composition \citep{2009ARA&A..47..481A};
Table 1 of \citet{2021arXiv210501661A}
summarises some of the differences with
the older generations of the 3D solar model.
Test calculations on an older version of the current 3D model
suggest that the differences in inferred abundance may be roughly around
$0.01\,\dex$ for most of the molecular indicators considered here.

For comparison purposes, as well as to estimate systematic errors,
several 1D hydrostatic models were also used in this study.
These are namely: a temporal- and horizontal-average 
(on surfaces of equal Rosseland mean optical depth)
of the full 3D model (Sect. 2.1.2 of \citealt{2018A&A...616A..89A}),
hereafter referred to as the \mtd{} model;
a theoretical \atmo{} model
(Appendix A of \citealt{2013A&A...557A..26M}), that is
calculated with the same radiative transfer scheme and input
physics as the 3D \stagger{} model;
the solar model from the theoretical \marcs{} grid
\citep{2008A&A...486..951G}, which is copiously used
in stellar abundance analyses;
and the semi-empirical model
of \citet{1974SoPh...39...19H}, hereafter the \handm{} model.
The latter model was originally 
constructed so as to reproduce a number of key 
observational diagnostics:
the centre-to-limb variation of the continuum
as well as the wings of Fraunhofer lines to probe
the $T\left(\tau\right)$ relation of the thicker layers of the photosphere;
and the centre-to-limb variations of the 
central intensities of around $900$ atomic and ionic lines
to probe the thinner layers \citep{1967ZA.....65..365H}.
We still refer to this last model because it has been used for many decades in a
very large number of works, 
but it should be noted that its temperature structure
may be slightly too warm in the
layers where typical atomic lines used for abundance purposes are formed 
(\citealt{2009ARA&A..47..481A}; see also \sect{discussion_oned}).

\subsection{Line formation calculations}
\label{method_lineformation}

The vertical intensities emergent from the 3D model,
and the four 1D models, were generated using the 3D LTE
radiative transfer code \scate{} \citep{2011A&A...529A.158H}.
The atomic and molecular partition functions and
equilibrium constants 
were updated to use the data from \citet{2016A&A...588A..96B}.
When computing the chemical  equilibrium,
\scate{} includes $\mathrm{H^{-}}$,  atoms and ions up to
and including the third ionisation stage (e.g. \ion{Fe}{III}),
as well as $12$ molecules of hydrogen, carbon, nitrogen, and oxygen:
$\mathrm{H_{2}}$, 
$\mathrm{H_{2}^{+}}$;
$\mathrm{C_{2}}$, 
$\mathrm{N_{2}}$, 
$\mathrm{O_{2}}$; 
$\mathrm{CH}$, 
$\mathrm{NH}$,
$\mathrm{OH}$;
$\mathrm{CN}$,
$\mathrm{CO}$,
$\mathrm{NO}$;
and $\mathrm{H_{2}O}$.

For the 3D model, the calculations were performed on $52$ snapshots
each roughly twenty minutes of solar time apart, to adequately sample
the solar granulation. Compared to the original simulation,
the horizontal resolution was reduced by a factor
of $3\times3$, from $240^{2}$ grid points to $80^{2}$ grid points,
while the vertical resolution was enhanced by a factor of $2$ 
near the optical surface.
The intensities were calculated without any extra 
microturbulent broadening, these non-thermal effects being
naturally accounted for due to the velocity and temperature inhomogeneities 
in the 3D model itself 
\citep{2000A&A...359..729A,2009LRSP....6....2N}.
The calculations on the 1D models were performed in almost the same
manner.  The main difference is that in the latter case, a 
depth-independent microturbulence was adopted,
$\vmic=1.0\,\kms$ \citep{1978A&A....70..537H}.

The abundances of all three of carbon, nitrogen, and oxygen
can influence the theoretical intensity depressions 
of all the lines in \tab{tab:linelist}, 
through changes to the equation of state and molecular balance.
Therefore, for all of the model atmospheres,
all line intensities were calculated for a three-dimensional rectilinear
grid of carbon, nitrogen, and oxygen abundances.
The abundances were taken to be 
between $-0.2\,\dex$ to $+0.2\,\dex$ in steps of $0.1\,\dex$ about central
values of $8.43$, $7.83$, and $8.69$, respectively 
\citep{2009ARA&A..47..481A},
resulting in a total of $5\times5\times5$ different 
chemical compositions.

The subsequent analysis 
is based on comparing theoretical
equivalent widths onto the observational data (\sect{method_observations}).
The theoretical equivalent widths were interpolated
in the three-dimensional abundance space using cubic splines.
For all of the lines predicted by 
the 3D model, the inferred elemental abundances 
all fell within the grids stipulated above.
However, for the 1D models, there were some deviations beyond
the edges of these grids, stipulated above.
Since the 1D results are not a priority of the present study,
linear extrapolations were applied in those cases.

\section{Abundance analysis and results}
\label{abundances}

\input{abundances}

\subsection{Preliminary comments}
\label{abundances_comments}

Combining the individual lines in \tab{tab:linelist} into 
a final estimate of the
elemental abundances is not a straightforward task.
One can see that, by taking an unweighted mean of all of the lines would
bias the result towards those groups or molecules
in \tab{tab:groups} having the largest number of analysed transitions. 
This is undesirable because lines of different molecules,
electronic systems, and $|\Delta\upnu|$ are susceptible to
different systematics, and the analysis should take advantage of this.
This is attempted here, as described in the remainder
of \sect{abundances}.

\subsection{Iterative procedure}
\label{abundances_analysis}

The analysis proceeds in an iterative fashion, 
because the inferred chemical composition,
and its uncertainty, has an effect on the 
theoretical equivalent widths and thus the inferred abundances.
In the initialisation step, 
a guess was made on the model-dependent abundances of
carbon, nitrogen, and oxygen, as well as on their uncertainties.
The choice of initial guess was not critical,
and varying it across the extent of the calculated grid
did not alter the final converged results.
For a given line, two of the three abundances
were fixed; the free parameter
is specified in \tab{tab:groups}. For CN and 
\comol{}, the free parameter is the less abundant 
element: nitrogen and carbon respectively.

In the first step, the lines were grouped according to
the rows in \tab{tab:groups}.
The assumption is
that lines of the same molecule, in the same electronic system,
with the same $|\Delta\upnu|$
have similar sensitivities
to systematic errors implicit in the model atmospheres
and molecular data, and should therefore be grouped together.
For each model, unweighted mean values,
$\lgeps{}\left(\mathrm{model; group}\right)$, were
determined by summing over all lines in a given group.
The squared uncertainties were calculated as follows:
\phantomsection\begin{IEEEeqnarray}{rCl}
\label{eq:grouperror}
    \upsigma^{2}\left(\mathrm{model; group}\right)&=& 
    \upsigma^{2}\left(\mathrm{group; atmos}\right)\IEEEnonumber \\
    &&+\upsigma^{2}\left(\mathrm{model; group; trend}\right)\IEEEnonumber \\
    &&+\upsigma^{2}\left(\mathrm{model; group; comp.}\right)\IEEEnonumber \\
    &&+\upsigma^{2}\left(\mathrm{model; group; stat.}\right)\, .
\end{IEEEeqnarray}
The first term in \eqn{eq:grouperror} accounts for the 
sensitivity of the group to the model atmosphere.
Here, this was estimated as
the standard error in the means of the results from the four different 
modelling paradigms: 3D, 
\mtd{}, 1D theoretical \atmo{}, and 1D semi-empirical \handm{};
it is therefore the same for all models.
The second term in \eqn{eq:grouperror} probes further shortcomings
that may be present in all model atmospheres,
for example systematic biases in the equivalent width measurements
or molecular data.
This was evaluated by fitting linear trends 
as functions of different line parameters
(namely, wavelength, excitation potential,
and logarithmic reduced equivalent width):
the most severe trend was identified, and half the difference
between the highest and lowest point of the fit was evaluated.
The third term in \eqn{eq:grouperror} accounts for the uncertainty in
the two fixed background abundances
propagated onto the fitted third abundance,
omitting the covariance terms.
The final term in \eqn{eq:grouperror}
gives the statistical uncertainty.  It was
evaluated as the sample standard deviation about the
fitted trend (noting that there are $N_{\text{line}}-2$ degrees of freedom),
divided by $\sqrt{N_{\text{line}}}$.
Any random errors in the equivalent widths and 
transition probabilities were assumed to be implicit
in this quantity.

In the second step, the results for different groups
were collapsed so as to
obtain mean values for each molecule and each model.
This was achieved via a weighted mean.
The weights were
calculated for each model and group, via the uncertainties 
evaluated using \eqn{eq:grouperror}:
\phantomsection\begin{IEEEeqnarray}{rCl}
\label{eq:speciesweights}
    \text{w}\left(\mathrm{model; group}\right) &=&
    \frac{\upsigma^{-2}\left(\mathrm{model; group}\right)}
    {\sum\upsigma^{-2}\left(\mathrm{model; group}\right)}\, .
\end{IEEEeqnarray}
Here, the summation in the denominator ensure that the
weights for all of the groups of a given molecule
sum up to unity (i.e.~that they are normalised).
The results for different groups, 
$\lgeps{}\left(\mathrm{model; group}\right)$,
were thus weighted
using \eqn{eq:speciesweights} and added together so as to determine
the weighted mean,
$\lgeps{}\left(\mathrm{model; molecule}\right)$.

To calculate the uncertainty in this weighted mean,
the four uncertainty components 
appearing in \eqn{eq:grouperror}
(atmospheric, trend, compositional,
and statistical, respectively) were individually propagated forward
using the following general formula:
\phantomsection\begin{IEEEeqnarray}{rCl}
\label{eq:error}
    \text{\textlangle}\sigma^{2}\text{\textrangle}&=&
    \sum_{i}^{n}w_{i}^{2}\sigma_{i}^{2}
    +\sum_{i}^{n}\sum_{j\neq
    i}^{n}w_{i}w_{j}\,\rho_{i,j}\,\sigma_{i}\sigma_{j}\, .
\end{IEEEeqnarray}
For example, for a given model and molecule,
the quantity 
$\upsigma\left(\mathrm{model; molecule; trend}\right)$
was determined by setting
$w_{i}$ and $w_{j}$ to $\text{w}\left(\mathrm{model; group}\right)$,
and $\sigma_{i}$ and $\sigma_{j}$
to $\upsigma\left(\mathrm{model; group; trend}\right)$,
and summing over the different groups as denoted by the indices $i$ and $j$ in 
\eqn{eq:error}.
The statistical and trend uncertainties were assumed
to be uncorrelated, and hence
the correlation coefficients $\rho_{i,j}$
were taken to be zero.
In contrast both the atmospheric and the compositional 
uncertainties were assumed to be systematic and thus 
perfectly correlated between different groups of the same 
molecule,
and therefore $\rho_{i,j}=1$.
For a given model and molecule,
the uncertainty,
$\upsigma\left(\mathrm{model; molecule}\right)$,
was then determined by adding these four propagated 
components together in quadrature.

In the third step, the results for different molecules
were combined into final determinations of the elemental abundances.
Similarly to the previous step,
these were calculated through a weighted mean,
summing over all molecules of a given element.
The four uncertainty components were propagated forward
again using \eqn{eq:error}, this time summing over
all molecules of a particular element,
rather than all groups of a particular molecule.
The final uncertainties in the elemental abundances 
were given by adding the
four components in quadrature.

With a new estimate of the elemental abundances in hand,
as well as an improved estimate of the composition error,
$\upsigma\left(\mathrm{model; group; comp.}\right)$, appearing 
in \eqn{eq:grouperror},
the next iteration cycle was initiated.
This proceeded until the elemental abundances had converged
to five decimal places.

One last note to make is that the line-by-line abundances
were adjusted to take into account
that the line formation calculations were performed
under the assumption that all CO is
in the major isotopologue, \comol{}.
Using for example Eq. 1 of \citet{2013ApJ...765...46A},
and the isotopic ratios advocated in 
\citet{2021arXiv210501661A},
this amounted to an increase in the carbon abundances
inferred from the \comol{} lines by only $0.006\,\dex$.

\subsection{Advocated elemental abundances}
\label{abundances_results}

We provide the converged line-by-line values
in the final five columns of \tab{tab:linelist},
corresponding to the five different models considered here.
In \tab{tab:abundances}, we show
the abundances inferred from the different groups, 
the collapsed results for the different
molecules, and our final advocated elemental abundances,
for each of the different models.
This table also shows the full breakdown of uncertainties, albeit
only for the 3D model, because in practice
they are dominated by the atmospheric uncertainty,
so are similar for the other models.

In \tab{tab:abundances},
the final advocated abundances as inferred from the molecules
are $\lgeps{C}=8.47\pm0.02$, 
$\lgeps{N}=7.89\pm0.04$, and $\lgeps{O}=8.70\pm0.04$.
The stipulated uncertainties are dominated 
by the atmospheric uncertainty.  These
were taken to be correlated ($\rho_{i,j}=1$ in \eqn{eq:error}),
which is why the individual groups and molecules
typically agree with the final values to better than these
stated precisions of $0.02$ to $0.04\,\dex$.
The largest deviation was found
for the NH $|\Delta\upnu|=1$ group,
for which the nitrogen abundance is found to be 
$0.03\,\dex$ larger than the final result.

\section{Discussion}
\label{discussion}

\begin{figure*}
    \begin{center}
        \includegraphics[scale=0.31]{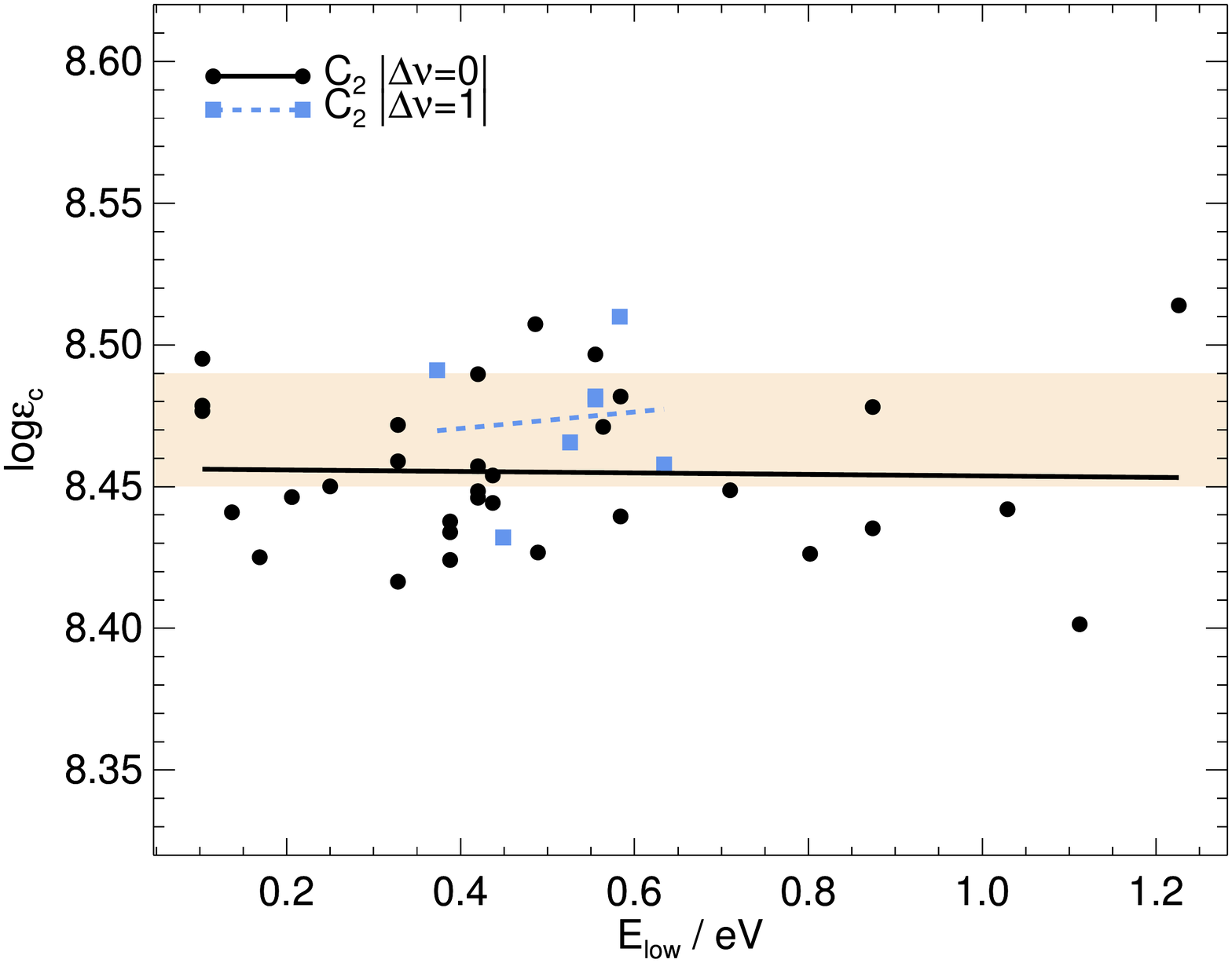}\includegraphics[scale=0.31]{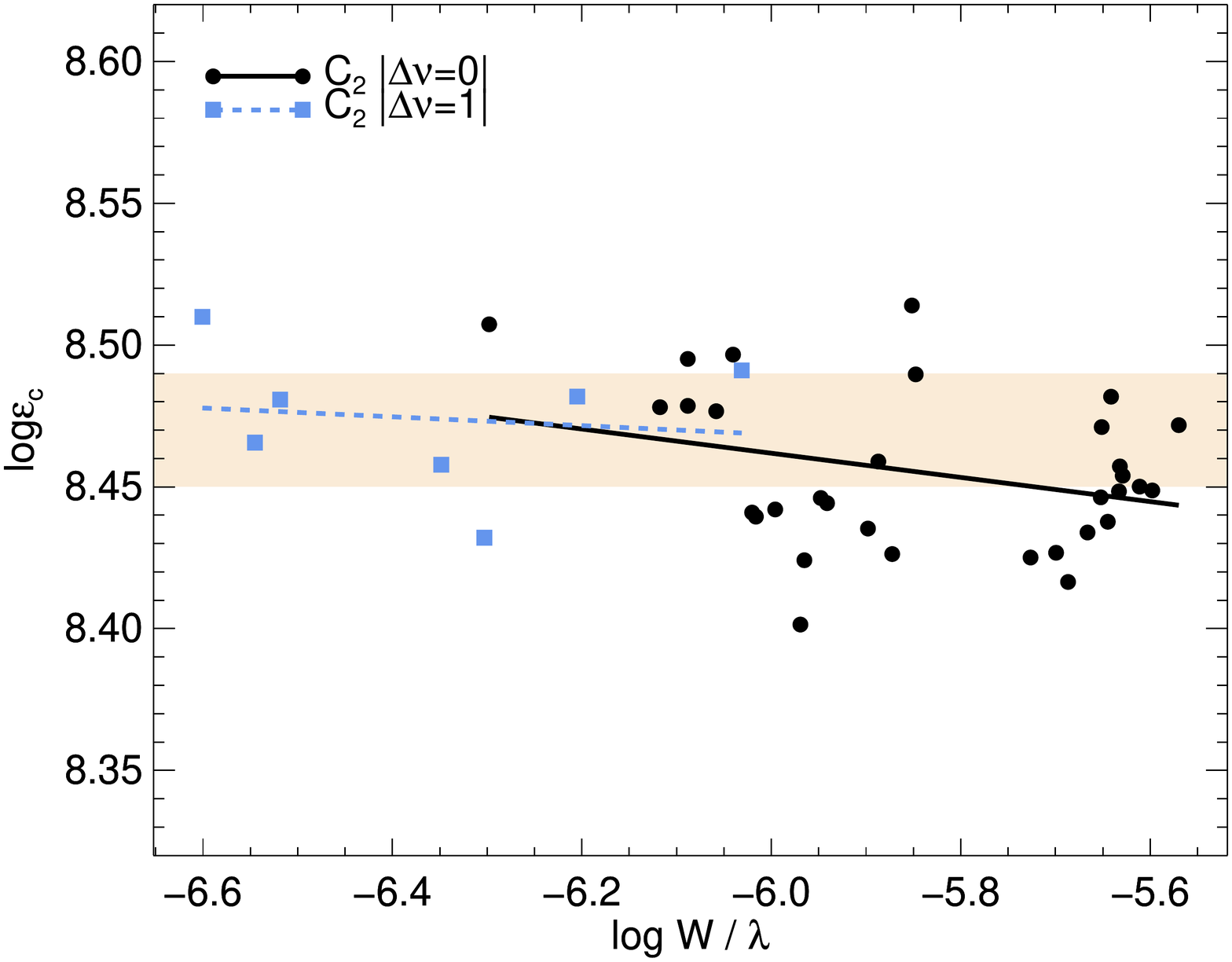}
        \includegraphics[scale=0.31]{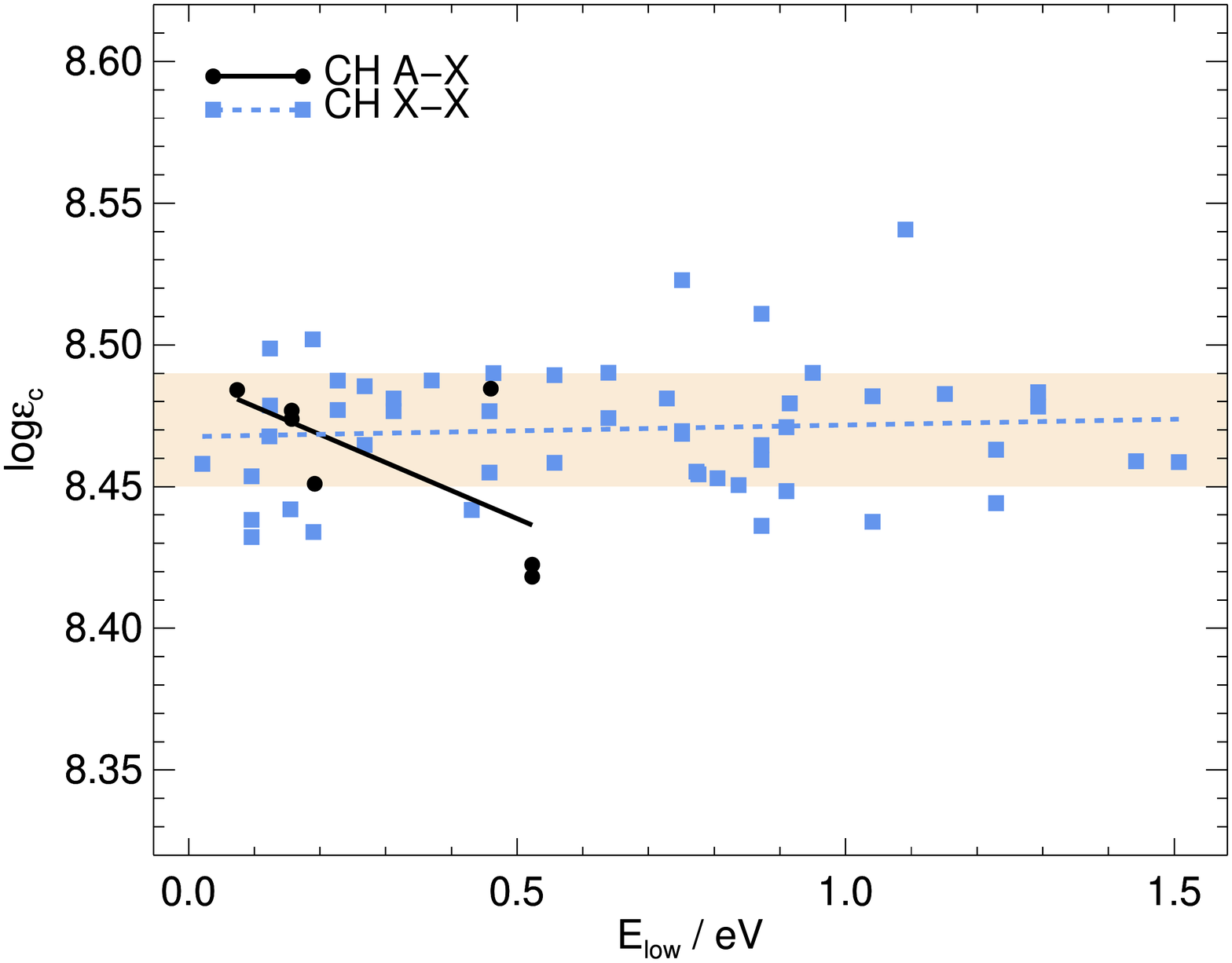}\includegraphics[scale=0.31]{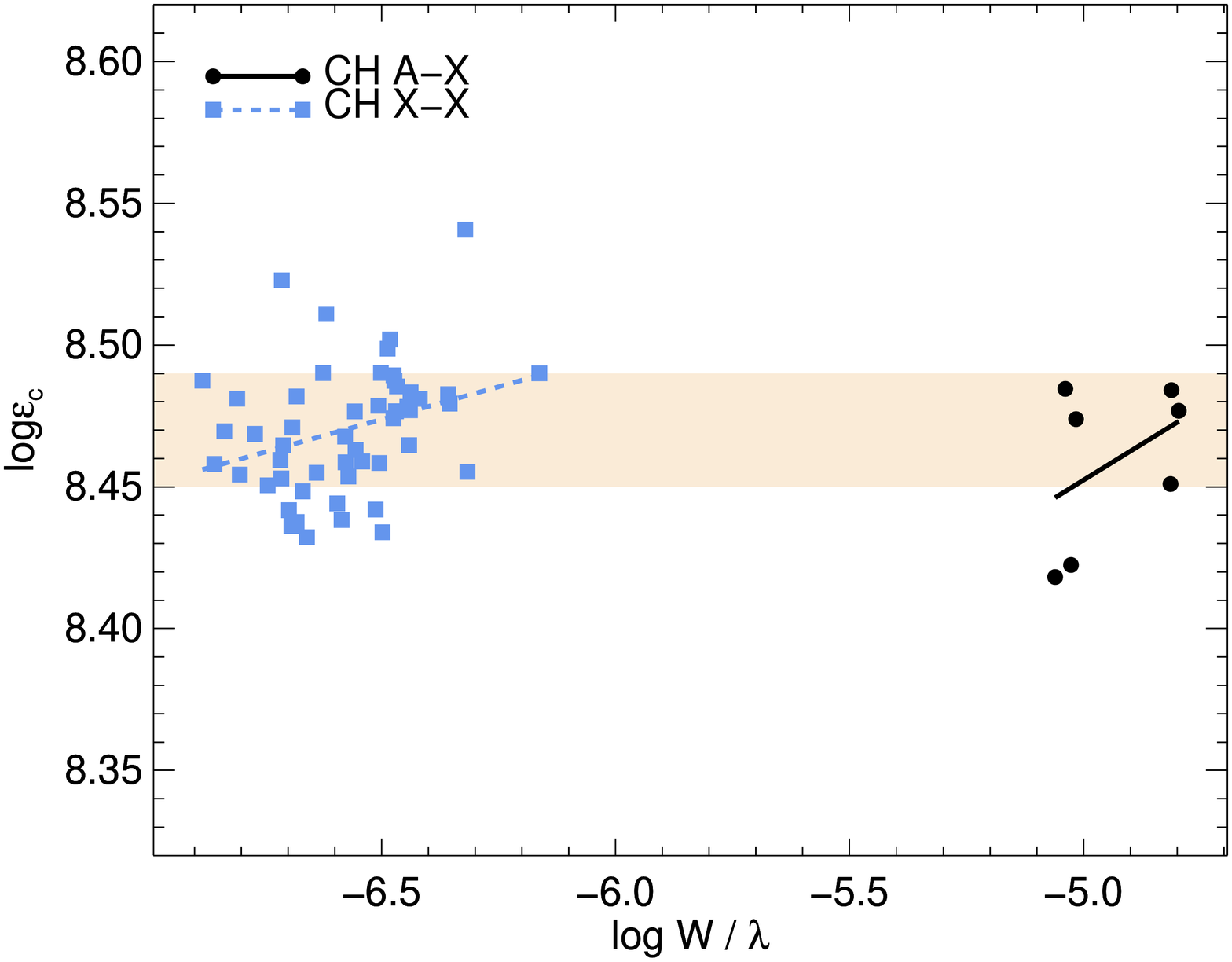}
        \includegraphics[scale=0.31]{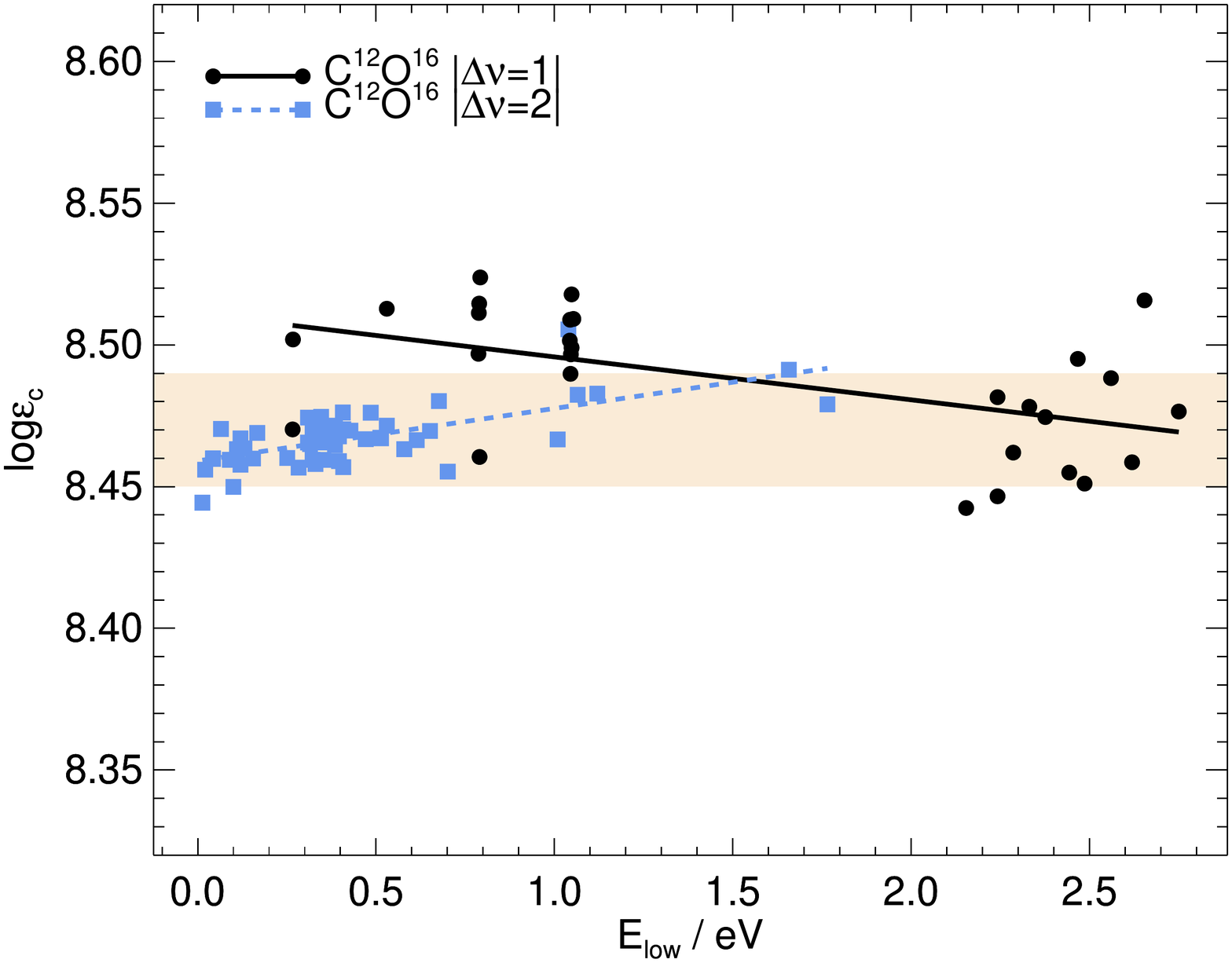}\includegraphics[scale=0.31]{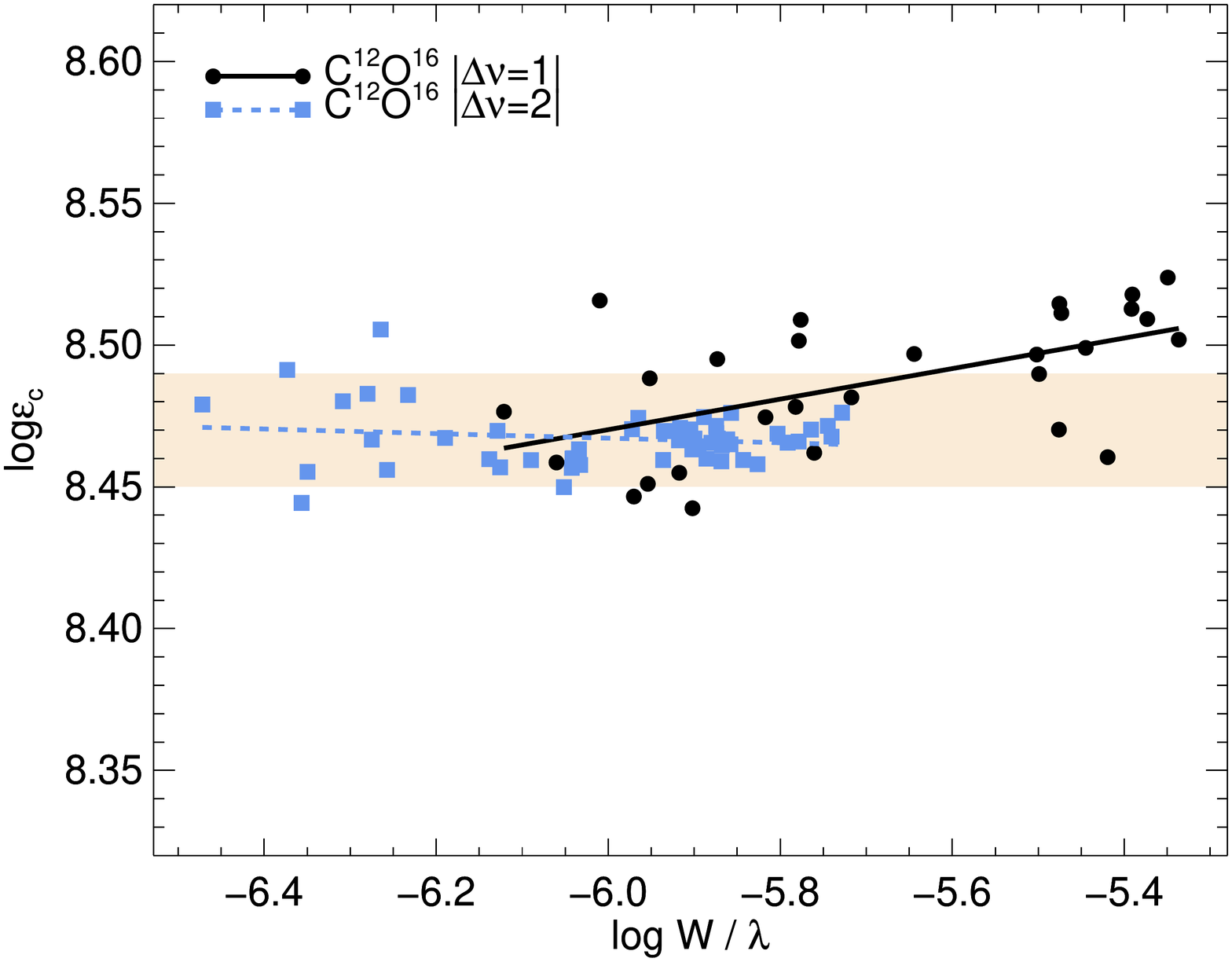}
        \caption{Carbon abundances inferred from 
        lines of \ctwo{}, CH, and \comol{} (rows), 
        against excitation potential and reduced
        equivalent width (columns).
        Linear regression lines have been overplotted. Shaded area 
        shows the advocated abundance and uncertainty,
        $\lgeps{C}=8.47\pm0.02$.}
        \label{fig:linescarbon}
    \end{center}
\end{figure*}

\begin{figure*}
    \begin{center}
        \includegraphics[scale=0.31]{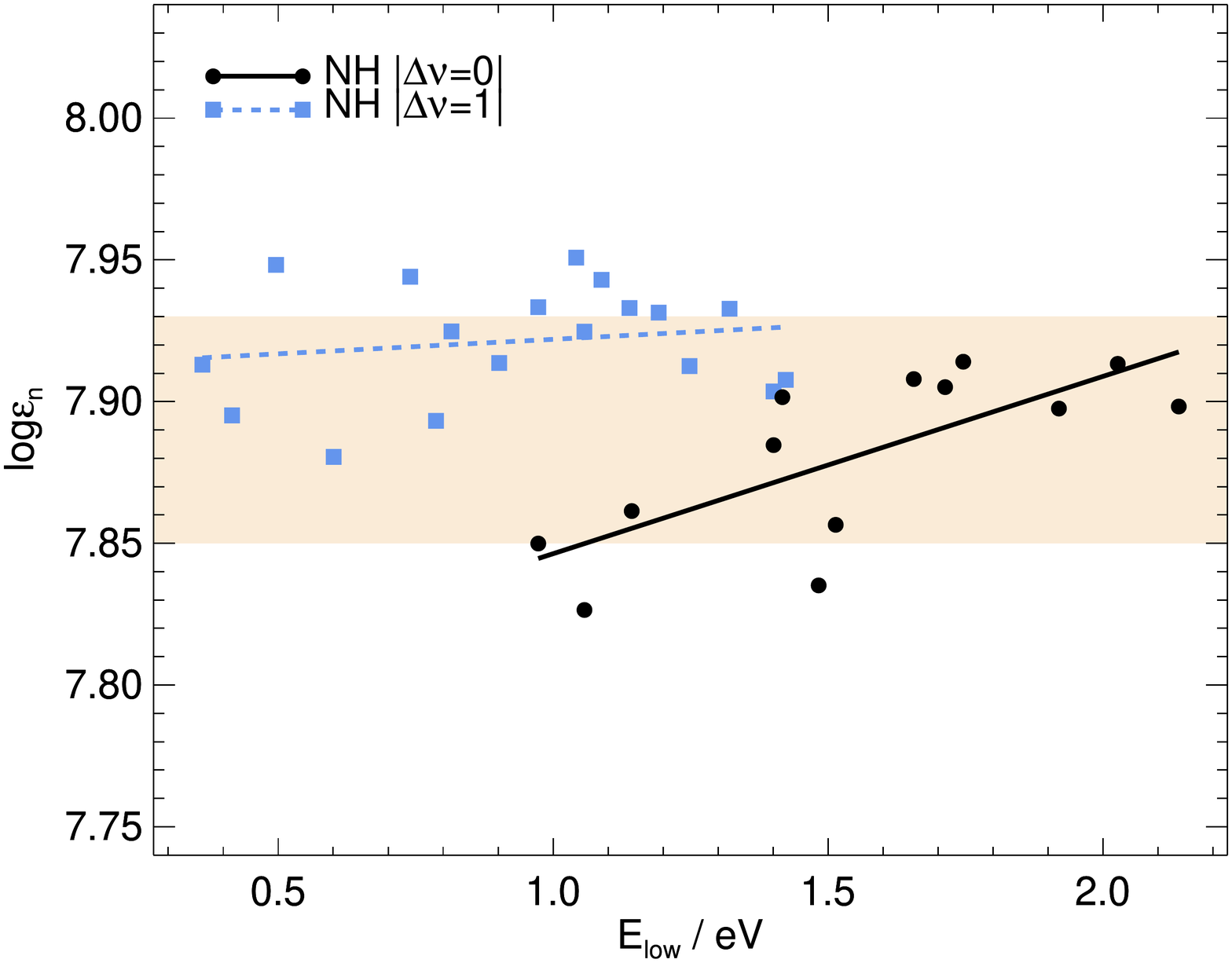}\includegraphics[scale=0.31]{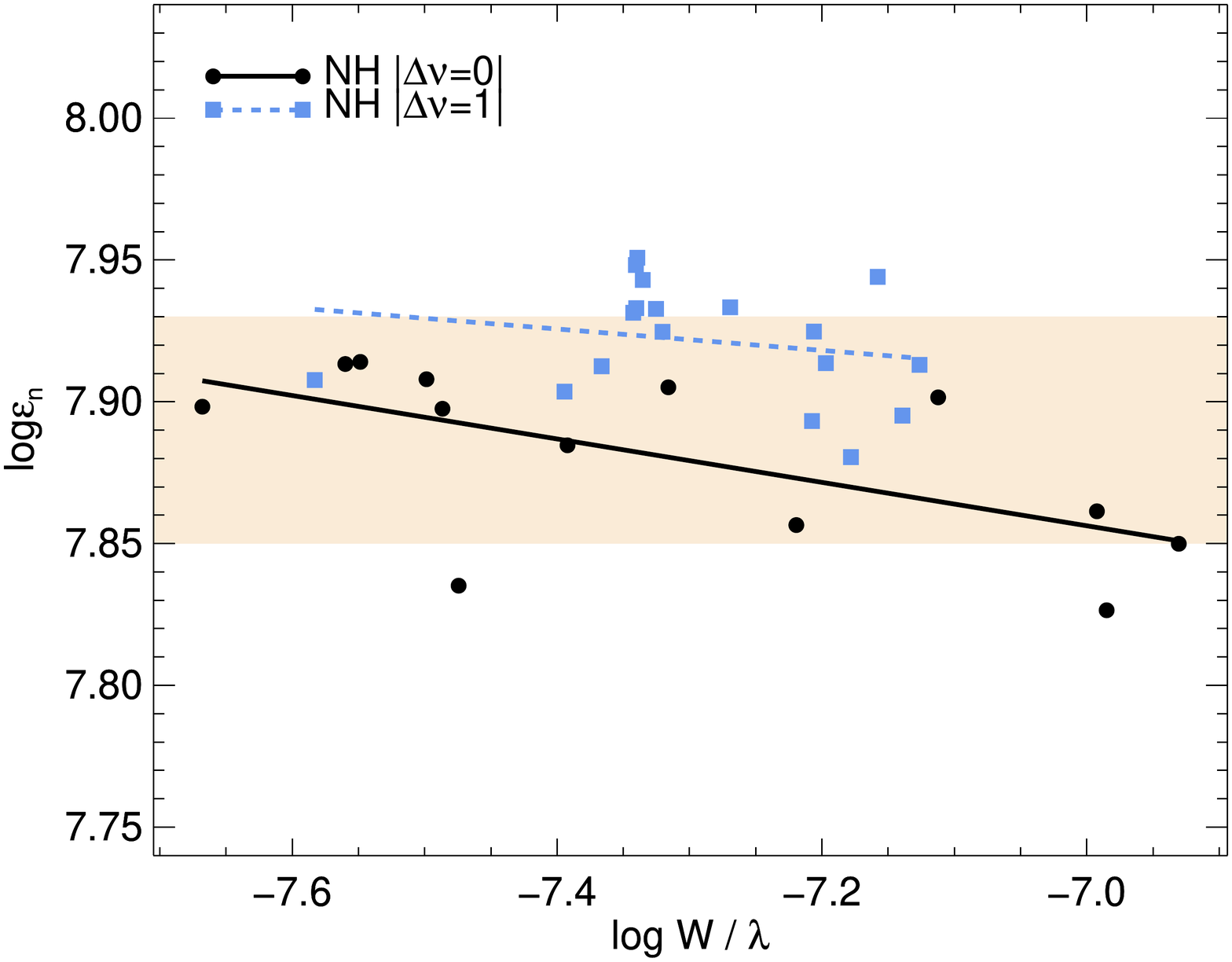}
        \includegraphics[scale=0.31]{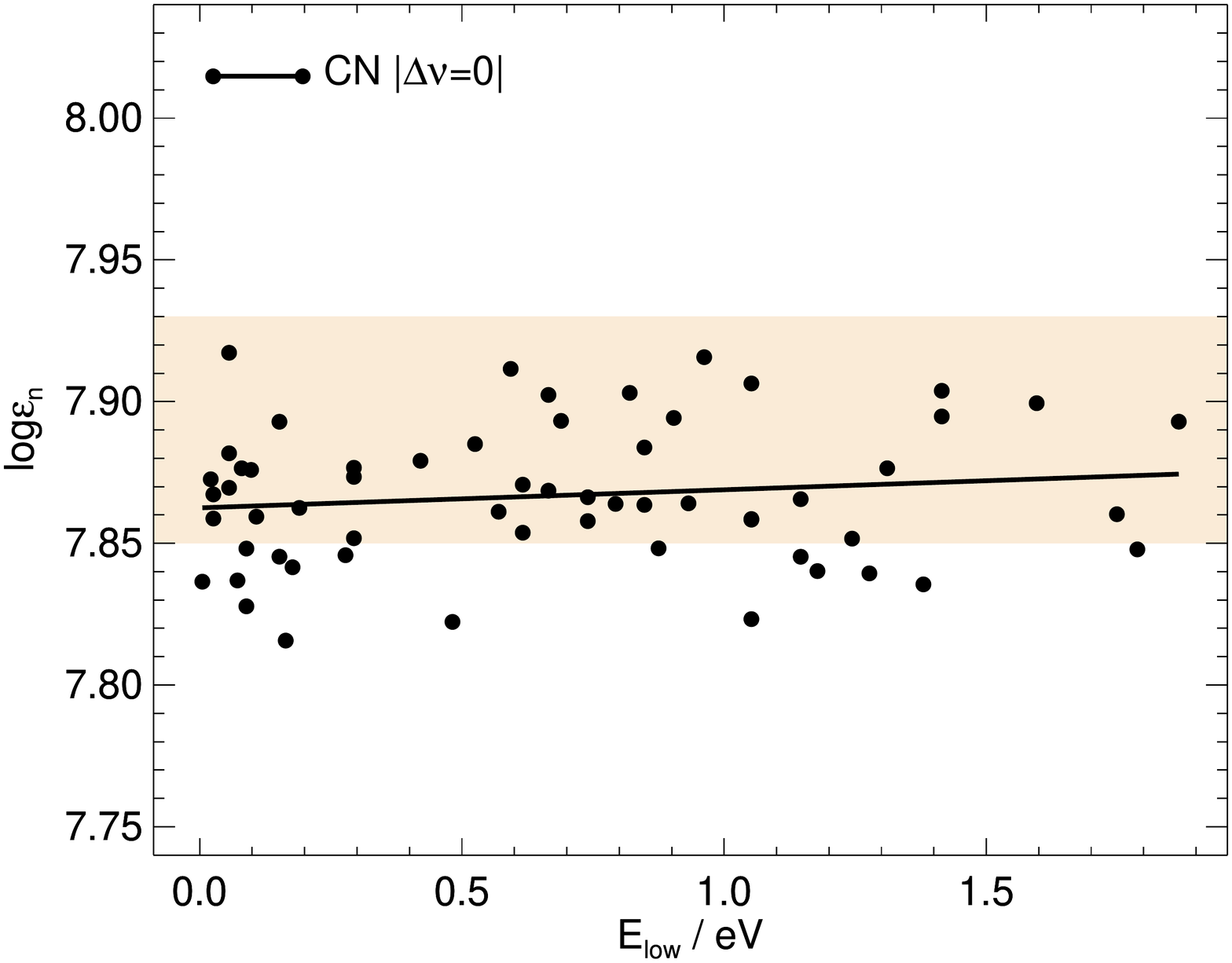}\includegraphics[scale=0.31]{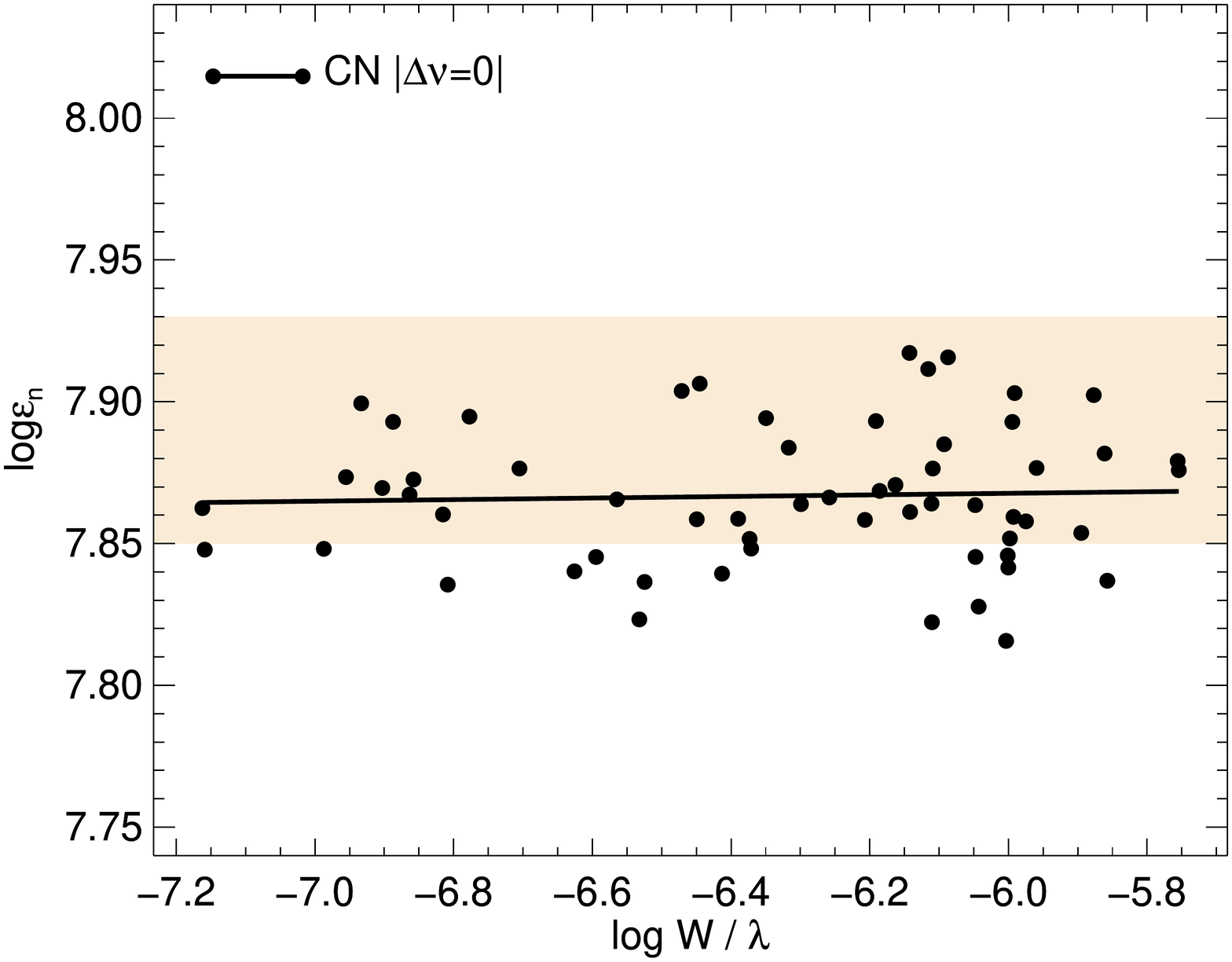}
        \caption{Nitrogen abundances inferred from 
        lines of NH and CN (rows), 
        against excitation potential and reduced
        equivalent width (columns).
        Linear regression lines have been overplotted. Shaded area 
        shows the advocated abundance and uncertainty,
        $\lgeps{N}=7.89\pm0.04$.}
        \label{fig:linesnitrogen}
    \end{center}
\end{figure*}

\begin{figure*}
    \begin{center}
        \includegraphics[scale=0.31]{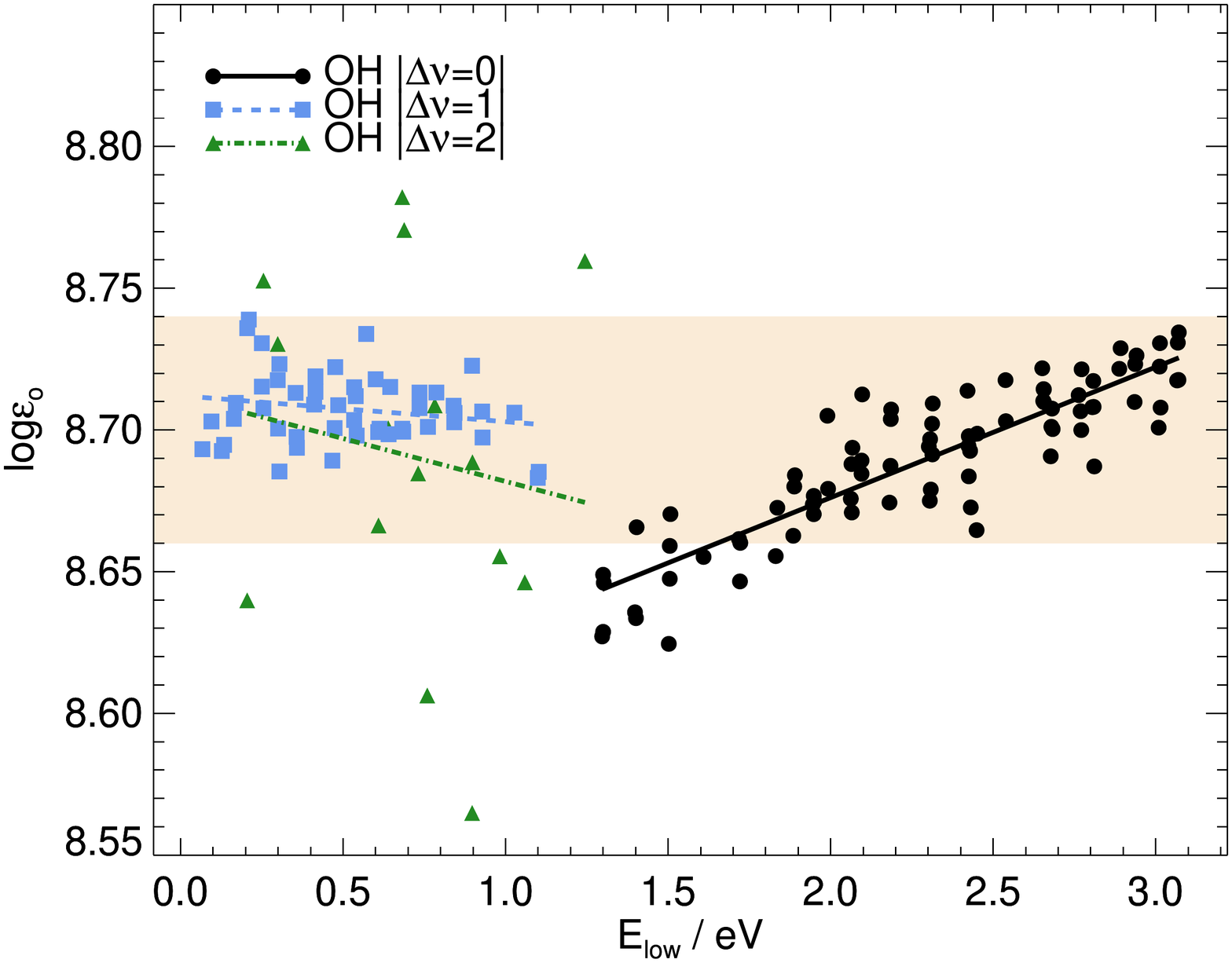}\includegraphics[scale=0.31]{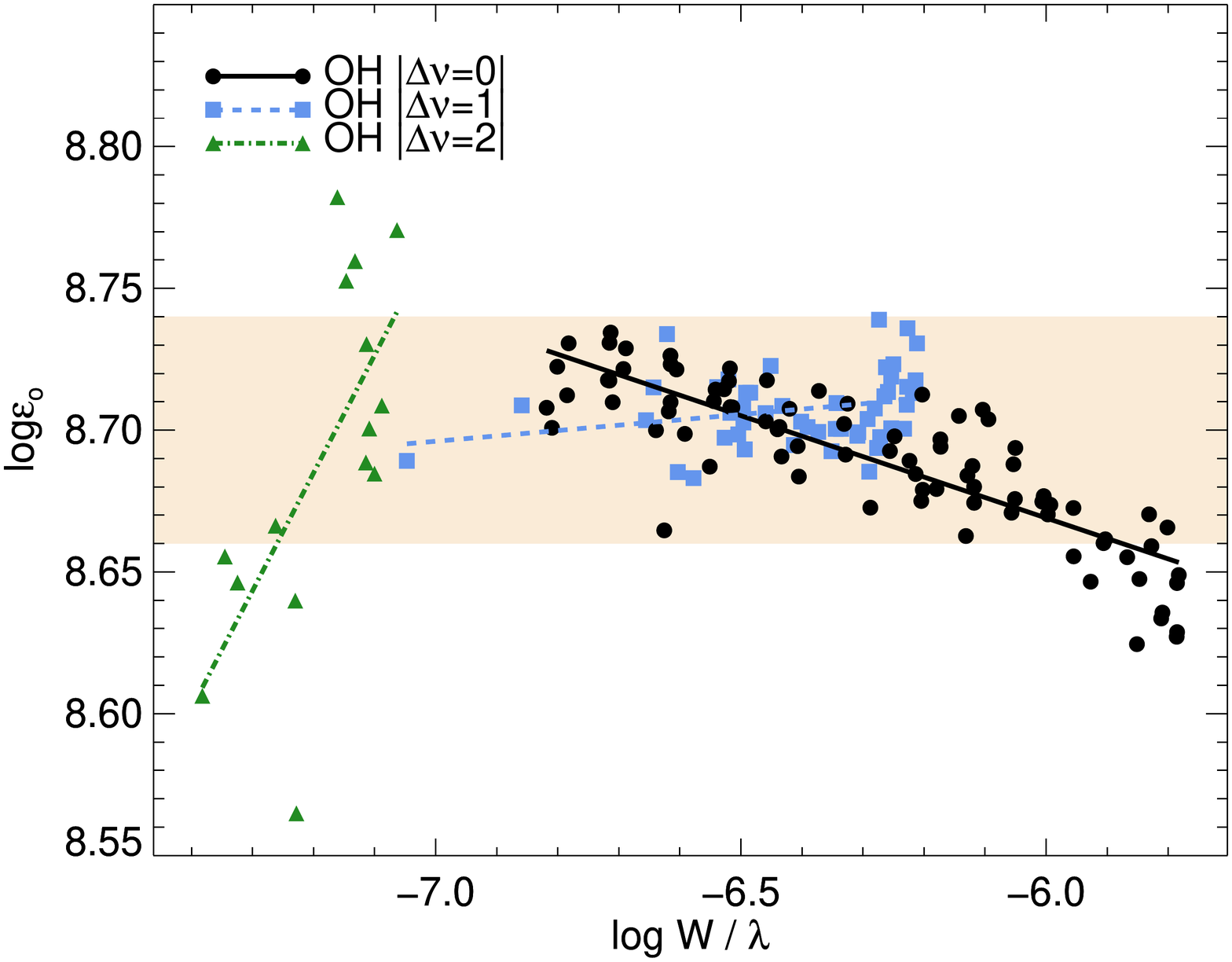}
        \caption{Oxygen abundances inferred from lines of OH, 
        against excitation potential and reduced
        equivalent width (columns).
        Linear regression lines have been overplotted. Shaded area 
        shows the advocated abundance and uncertainty,
        $\lgeps{O}=8.70\pm0.04$.}
        \label{fig:linesoxygen}
    \end{center}
\end{figure*}

\begin{figure*}
    \begin{center}
        \includegraphics[scale=0.6]{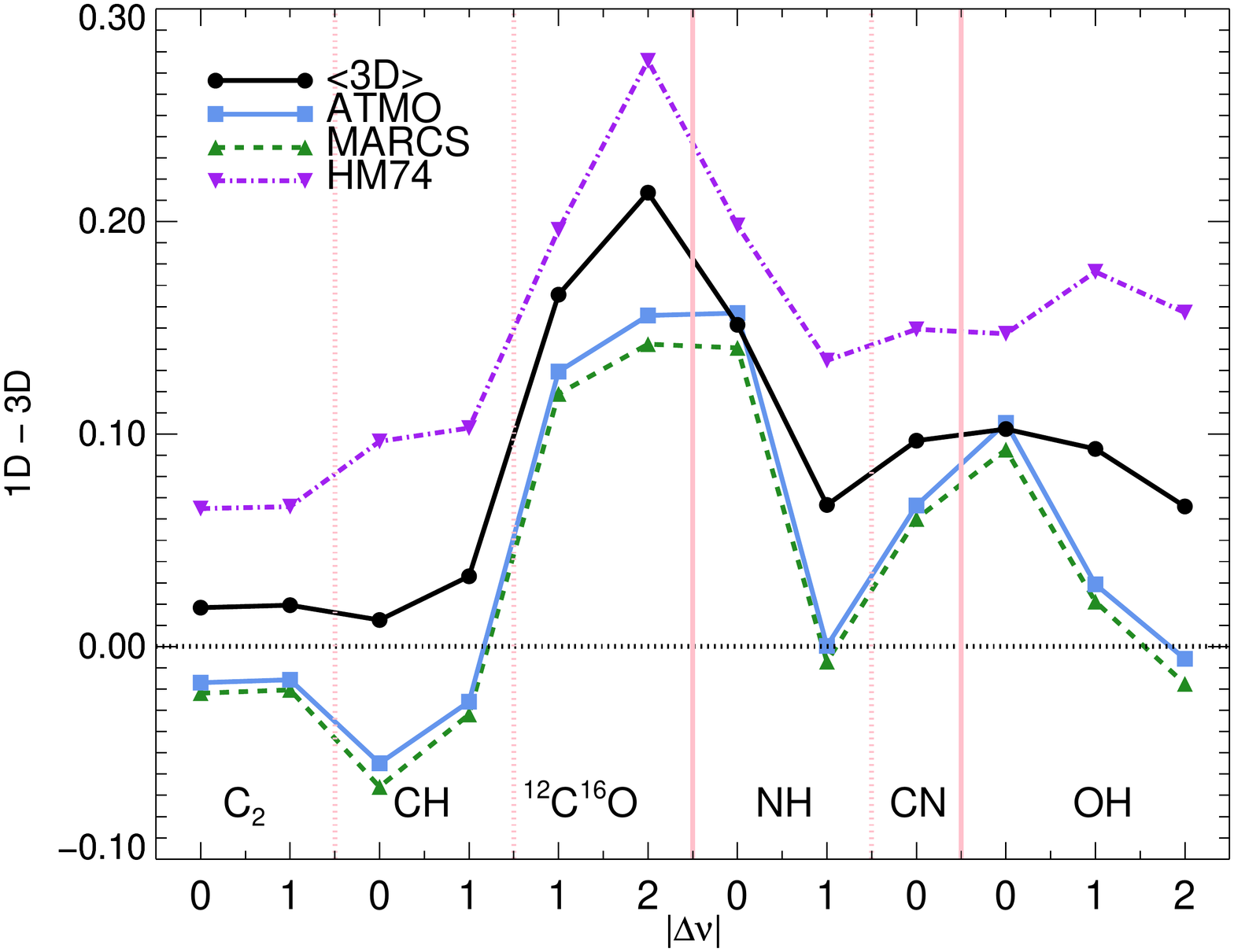}
        \caption{Difference between 1D and 3D elemental
        abundances for the different groups in \tab{tab:groups}.}
        \label{fig:compare}
    \end{center}
\end{figure*}

\subsection{Line-by-line results in the 3D model}
\label{discussion_trends}

Figs.~\ref{fig:linescarbon}, \ref{fig:linesnitrogen},
and \ref{fig:linesoxygen} illustrate
the line-by-line abundances for the 3D model,
for carbon, nitrogen, and oxygen respectively. 
The individual panels show the results for different 
molecules,
plotted against excitation potential and reduced equivalent width. 
In addition, for each group of indicators, the difference between the 
average results from the 1D models and from the 3D model
are shown in \fig{fig:compare}.

The line-by-line standard deviations,
i.e.~the scatter about the mean trends in
Figs.~\ref{fig:linescarbon}, \ref{fig:linesnitrogen},
and \ref{fig:linesoxygen}, are found to be
around $0.02\,\dex$.  The lowest scatters are
found for the \comol{} $|\Delta\upnu|=2$ and 
OH $|\Delta\upnu|=1$ groups
($0.01\,\dex$).  
This scatter can be partially explained by random measurement errors,
that can reach of the order $2.5\%$ or
$0.01\,\dex$ (\sect{method_observations});
random uncertainties in the adopted transition probabilities
likely also contribute.
An anomalously large scatter is found for
the OH $|\Delta\upnu|=2$ group ($0.05\,\dex$).
This group displays a severe trend with 
reduced equivalent width, which likely reflects particular
difficulties in measuring these extremely weak 
lines to high precision and high accuracy.

The two rotational groups, NH $|\Delta\upnu|=0$ and
OH $|\Delta\upnu|=0$, show rather strong trends
in \fig{fig:linesnitrogen} and \fig{fig:linesoxygen}.
These lines are in the far infrared, 
between $11300$ and $15000\,\nm$ for NH 
and between $9400$ and $12300\,\nm$ for OH;
i.e; further to the IR than the
rovibrational lines of CH, NH and OH (\tab{tab:groups}). 
Consequently, these infrared rotational lines form 
higher up in the atmosphere
than the rovibrational lines
(at Rosseland mean optical depths that are nearly $1\,\dex$ smaller),
on account of the strong wavelength dependence of the
dominant H$^-$ free-free continuous opacity.  
Thus the depicted trends may in part
reflect inadequacies in the uppermost layers of the 3D model due to for example
the lack of magnetic fields, and absence of a chromosphere.
The OH $|\Delta\upnu|=0$ lines are stronger 
than the NH $|\Delta\upnu|=0$ lines
(median equivalent widths of $5.6\,\mathrm{pm}$
and $1.7\,\mathrm{pm}$, respectively),
and typically form at greater heights; accordingly,
they display a more prominent trend relative to the scatter. 
In both cases, the results from the deeper-forming fainter lines 
tend to agree better with the rovibrational lines, as expected.

The figures provide some insight on the most and least reliable
abundance indicators, for each molecule, and for each element.
For carbon, the \ctwo{} Swan bands emerge as the most reliable
indicators.  They are absent of significant trends
with line parameters (\fig{fig:linescarbon}),
and also have low sensitivity to the 
model atmosphere (\fig{fig:compare}).
The \comol{} lines are the least reliable ones,
despite their low scatter.
This is expected because of the temperature sensitivity
of their formation, and thus their sensitivity to the 
model atmosphere as is reflected 
in \fig{fig:compare}.
For nitrogen, NH and CN are found to have similar weight.
Within NH, the 
$|\Delta\upnu|=1$ lines have the most weight, as the 
$|\Delta\upnu|=0$ lines show more pronounced trends as discussed above.
Finally for oxygen all three groups have similar
sensitivities to the model atmosphere;
the OH $|\Delta\upnu|=1$ lines are the most
reliable, because of the noticeable trends in
the OH $|\Delta\upnu|=0$ and OH $|\Delta\upnu|=2$ groups.

\subsection{Comparison with the 1D models}
\label{discussion_oned}

Typically, the results from the various 1D models 
are larger than those from the 3D model --- in other words,
the 3D effects are usually negative.
For carbon, nitrogen, and oxygen respectively,
the \mtd{} results are $0.02$, $0.10$, and $0.10\,\dex$ larger;
the \atmo{} and \marcs{} 
results are $-0.02$, $0.06$, and $0.06\,\dex$ larger,
and the \handm{}
results are $0.08$, $0.15$, and $0.15\,\dex$ larger.
This systematic effect is
because of the enhanced formation of molecules
in the cool pockets of gas that arise due to solar convection.
It is particularly effective for the
temperature-sensitive \comol{} lines, as well as the NH $|\Delta\nu|=0$ lines
and to a lesser extent the OH $|\Delta\nu|=0$ lines
(\fig{fig:compare}) that form in the uppermost regions of the photosphere.

It is perhaps counter-intuitive that the 
results from the 
semi-empirical \handm{} model deviates quite strongly from
the 3D model. This is related to the smaller temperature
gradient of that model, with temperatures
around $50\,\K$ larger in the line-forming region relative to the 
\mtd{} model (\fig{fig:atmos}), 
as previously discussed by \citet{2009ARA&A..47..481A}.
While the origin of this temperature difference is not completely clear,
we can speculate on a few possible reasons for it.
The \handm{} model presented in 
\citet{1974SoPh...39...19H} is an updated version
of \citet{1967ZA.....65..365H}, with the photospheric temperature
stratification in the line-forming region determined primarily by fitting the
centre-to-limb variations of the central intensities of some $900$
atomic and ionic lines, under the assumption of LTE.
Our measurements suggest that the line
depths given by \citet{1967ZA.....65..365H} are slightly
but systematically smaller than those in
in the more recent and higher-quality Li\`ege and Hamburg 
solar atlases employed here.
This could be associated with difficulties 
in correcting for stray light and the lower spectral resolution 
of the spectrograph used by \citet{1967ZA.....65..365H}, 
although systematic differences
in the continuum normalisation
could also play a role.  Adopting shallower line depths in the
construction of the \handm{} model could at least
partially explain the shallower temperature gradient and thus
the larger estimate of the temperature in the line forming regions. The
assumption of LTE for neutral species of low to moderate ionisation potential
including \ion{Mg}{I} and \ion{Fe}{I} may lead to a similar effect 
(non-LTE masking; \citealt{1982A&A...115..104R});
although, this is compensated by their choice of
high elemental abundances $\lgeps{Mg}=\lgeps{Fe}=7.60$
when determining the electron number densities. It can be noted that 
newer semi-empirical solar models of the quiet Sun such as those 
by \citet{1999A&A...347..348G}
and \citet{2001ApJ...558..830A}, have temperature structures
more closely resembling that of the \mtd{} model.

\subsection{Comparison with earlier studies of molecules}
\label{discussion_oldmolecules}

The abundance values advocated in \sect{abundances_results} generally
agree well with those of \citet{2009ARA&A..47..481A},
which was an updated analysis of \citet{2005A&A...431..693A}
for carbon and \citet{2004A&A...417..751A} for oxygen using an improved 3D solar model. 
There are, nevertheless, some puzzling discrepancies:
the most prominent ones are for CH and NH,
which are around $0.03$ to $0.05\,\dex$ larger here.
This offset does not seem to be due mainly to the model atmosphere 
(being also apparent for the results from the 1D \marcs{} models), 
transition probabilities, nor the adopted equivalent widths.
We speculate that incremental updates to the line formation code,
and equation of state solver 
(including input data such as 
partition functions and equilibrium constants)
could be the cause of the offset.
While the average result for OH is consistent, it
should be noted that
the line-by-line results here do not show
the strong non-linear trends apparent in 
Fig.~10 of \citet{2004A&A...417..751A}.

The \comol{} lines were not used to inform the
carbon or oxygen abundances in  \citet{2009ARA&A..47..481A}.
This is primarily because of their temperature sensitivity,
although they may also be sensitive to departures from 
instantaneous chemical equilibrium
\citep{2000ApJ...536..481U, 2003ApJ...588L..61A};
such effects are neglected in the present study.
Nevertheless, elemental abundances via 
\comol{} lines were also derived by \citet{2006A&A...456..675S} 
(with the same 3D solar model as in \citealt{2005A&A...431..693A}).
Fixing $\lgeps{O}=8.66$ they obtained $\lgeps{C}=8.48$ and $8.40$
from \comol{} $|\Delta\upnu|=1$ lines of
low and high excitation potential, respectively;
and $\lgeps{C}=8.37$ from \comol{} $|\Delta\upnu|=2$ lines.
The line selection adopted here derives from theirs,
albeit with updated transition probabilities 
from \citet{2015ApJS..216...15L} that are much improved as discussed in
\citet{2018NatCo...9..908L}.
We find consistent results from all of these 
\comol{} lines of around $\lgeps{C}=8.48\,\dex$,
as illustrated in \fig{fig:linescarbon}.
Test calculations suggest that 
the updates to the 3D model help to
ensure congruous results from the different \comol{} lines.

\subsection{Comparison with atomic results}
\label{discussion_atoms}

It is illuminating to compare the molecular results derived here
with atomic results present in the literature.
Recently, 3D non-LTE studies of \ion{C}{I} \citep{2019A&A...624A.111A,
2021MNRAS.502.3780L},
\ion{N}{I} \citep{2020A&A...636A.120A}, 
and \ion{O}{I} \citep{2018A&A...616A..89A} lines
have been carried out. These studies are based on
the same model atmosphere as that used here.
In addition, 3D LTE studies with 1D non-LTE corrections
have been carried out for these species by 
\citet{2008A&A...488.1031C,2009A&A...498..877C,2010A&A...514A..92C,
2013A&A...554A.126C,2015A&A...579A..88C},
as well as full 3D non-LTE calculations of the 
\ion{O}{I} $777\,\nm$ triplet by \citet{2015A&A...583A..57S}
on an independent 3D solar model computed with the \cobold{} code \citep{2012JCoPh.231..919F}.

A comparison can be found in 
\citet{2021arXiv210501661A}; the main findings are summarised here. 
Due to several different updates,
in particular a systematic shift in the transition probabilities
for \ion{C}{I} \citep{2021MNRAS.502.3780L},
the various carbon abundance indicators all now
point to a value of around $8.46\,\dex$,
which is $0.03\,\dex$ larger than the value advocated by
\citet{2009ARA&A..47..481A}.
For oxygen, the agreement is also very satisfactory,
with the [\ion{O}{I}] lines and 
\ion{O}{I} lines indicating on average
slightly higher and lower results by around $0.02\,\dex$,
compared to the molecular result of $8.70\,\dex$ found here.

For nitrogen, however, there is a prominent discrepancy 
with the atomic lines.  With the present 
3D model, the \ion{N}{I} lines give $\lgeps{N}=7.77\pm0.05$.
This is
significantly lower than the present molecular result
of $7.89\pm0.04\,\dex$;
a difference of $0.12\,\dex$.
The origin of this problem is unclear.
Systematic errors in the transition probabilities
or other molecular data cannot be ruled out, but seem unlikely
as discussed in \citet{2021arXiv210501661A}.
The atomic and molecular lines have opposite temperature
sensitivities; consequently,
errors in the temperature structure of the 3D model
could help explain the discrepancy,
although the large changes needed
would worsen the agreement between the various carbon 
and oxygen abundance
indicators, as well as the excellent agreement 
with various other solar observables
as studied by \citet{2013A&A...554A.118P} for the previous
generation of the 3D \stagger{} model.
If this is indeed the case, the 
true solar nitrogen abundance would be close to the mean result
from the atoms and molecules: $\lgeps{N}=7.83$, 
as advocated in \citet{2021arXiv210501661A}.
It can be noted that the various
1D models show even larger discrepancies:
for instance, the \handm{} model indicates a difference of $0.18\,\dex$, with
$\lgeps{N}=7.86$ from the \ion{N}{I} lines 
(unpublished results from \citealt{2020A&A...636A.120A}),
and $\lgeps{N}=8.04$ from the molecules (\tab{tab:abundances}).

More recently, \citet{2021MNRAS.tmp.1964B} have 
studied of the \ion{O}{I} $777\,\nm$ triplet lines
and the [\ion{O}{I}] $630\,\nm$ line, using
13 snapshots of the 3D model solar atmosphere
from the \stagger{}-grid
\citep{2011JPhCS.328a2003C,2013A&A...557A..26M}.
This 3D model is similar to that employed
in \citet{2018A&A...616A..89A,2019A&A...624A.111A,2020A&A...636A.120A}
as well as in the present study:
it was computed with the \stagger{} code, employing
the standard solar chemical composition of \citet{2009ARA&A..47..481A};
and the 13 snapshots have a mean effective temperature of $5773\,\K$.
The authors employed column-by-column non-LTE radiative transfer
(the so-called 1.5D non-LTE approach);
Fig.~6 of that paper suggests that, at least in this particular case,
this approximation 
corresponds to errors that are much less than $0.01\,\dex$,
compared to the more realistic 3D non-LTE approach.
From the \ion{O}{I} and [\ion{O}{I}] features
they obtained $\lgeps{O}=8.74\pm0.03$ and 
$\lgeps{O}=8.77\pm0.05$, respectively,
finally advocating $\lgeps{O}=8.75\pm0.03$.
This is slightly larger than one standard deviation 
away from our own measurements from the same features:
$8.69\pm0.03\,\dex$ and $8.70\pm0.05\,\dex$ respectively
\citep{2018A&A...616A..89A,2021arXiv210501661A}.
Given the importance of the solar oxygen abundance
on the solar modelling problem, we briefly
discuss below from where these differences may originate.

For the \ion{O}{I} $777\,\nm$ triplet lines, 
$0.026\,\dex$ of the difference
can be explained by the choice of transition probabilities.
\citet{2021MNRAS.tmp.1964B} adopted the recent values
of \citet{2018ApJS..239...11C} that are based on
quantum defect theory \citep[e.g.][]{1966PPS....88..801S},
averaged together with those 
of \citet{1991JPhB...24.3943H}
that are based on the (more sophisticated)
configuration interaction (CI) approach
(as implemented in the CIV3 code; \citealt{1975CoPhC...9..141H}).
In contrast, \citet{2018A&A...616A..89A} used only the latter data set.
For the [\ion{O}{I}] $630\,\nm$ line, most of the discrepancy
probably reflects the inconsistent treatment of the \ion{N}{I} blend.
\citet{2021MNRAS.tmp.1964B} assumed 
$\lgeps{Ni}=6.23$ \citep{2003ApJ...591.1220L}, and subsequently
corrected the strength of the blend for absolute 1.5D non-LTE effects,
rather than deriving a nickel abundance consistent with their model atmosphere
and radiative transfer approach using other, clean
\ion{Ni}{I} lines (see the discussion in \citealt{2009ApJ...691L.119S}).
Effectively, \citet{2021MNRAS.tmp.1964B} adopted a 
$0.05\,\dex$ smaller nickel contribution to the $630\,\nm$ feature 
compared to \citet{2021arXiv210501661A},
which translates to a $0.03\,\dex$ larger inferred oxygen abundance.
The remaining discrepancies may be due to differences between their new FTS
data, and the Li\`ege and Hamburg solar atlases employed in
\citet{2021arXiv210501661A}.  They may also be partly due to systematic 
differences in the 3D non-LTE radiative transfer codes employed.

\subsection{Advocated solar elemental abundances and metallicity}
\label{discussion_final}

It is preferable to fold in both
atomic and molecular results to obtain final estimates
on the elemental abundances, in particular 
for nitrogen.
These advocated values can be found in \citet{2021arXiv210501661A}:
$\lgeps{C}=8.46$, $\lgeps{N}=7.83$, $\lgeps{O}=8.69$.
Nevertheless, it is illuminating to consider the impact
on the solar metallicity 
when adopting the carbon, nitrogen, and oxygen abundances
inferred here, these values being slightly larger than 
the atom-molecule averages.
With the chemical mixture of \citet{2021arXiv210501661A}, the
metal mass fraction of the solar surface 
is $Z=0.0139\pm0.0006$; or, relative to the hydrogen mass fraction,
$Z/X=0.0187\pm0.0009$.
With the molecular abundances determined here, 
these values increase to  
$Z=0.0142$ and $Z/X=0.0191$, 
where $45\%$ of the difference is due to the $0.01\,\dex$
increase in oxygen abundance,
$35\%$ is due to the relatively large $0.06\,\dex$ increase in
nitrogen abundance, and just $20\%$ is due to the $0.01\,\dex$
increase in carbon abundance.
The differences thus remain well within the stipulated uncertainties.

\section{Conclusion}
\label{conclusion}

We have analysed $408$ molecular lines of carbon, nitrogen, and oxygen
in the solar intensity spectrum 
using LTE line formation calculations in a
3D radiative-hydrodynamic simulation of the photosphere.
From this analysis, advocated elemental abundances 
of $\lgeps{C}=8.47\pm0.02$, 
$\lgeps{N}=7.89\pm0.04$, and $\lgeps{O}=8.70\pm0.04$
were obtained. For carbon and oxygen,
these are in good agreement with earlier 
3D non-LTE analyses of atomic diagnostics 
\citep{2018A&A...616A..89A,2019A&A...624A.111A,
2021MNRAS.502.3780L,2021arXiv210501661A}. 
However, for nitrogen there is a discrepancy
approaching $3\sigma$ of $0.12\,\dex$, 
compared to the results presented in
\citet{2020A&A...636A.120A}.

The elemental abundances of these elements remain
relevant in the solar modelling problem.
Although oxygen dominates the opacity in the most discrepant region,
carbon and nitrogen are also relevant because
they are all mainly present as neutral atoms in the solar photosphere,
their molecules are closely coupled together 
via \comol{} and CN, and they are similarly depleted in meteorites
\citep{2009LanB...4B...44L,2014pacs.book...15P} and must
therefore be inferred via spectroscopic methods.
The solar oxygen abundance determined here, consistent
with the atomic results, is just $0.01\,\dex$ larger
than that advocated in \citet{2009ARA&A..47..481A}
and unfortunately does little to alleviate the 
solar modelling problem.  The carbon and 
nitrogen abundances determined here are
$0.03$ and $0.06\,\dex$ larger than the 
values advocated by \citet{2009ARA&A..47..481A}.
Although this may only slightly reduce the discrepancy
with helioseismology, this revision does indicate that it
may be worthwhile to continue improving the atomic and molecular
data as well as the model atmospheres and line formation methods.

\begin{acknowledgements}
We are grateful to
Matthias Steffen for providing a detailed and insightful
referee report, which led to a number of improvements to this manuscript.
We thank T.~Masseron and C.~Sneden for their assistance with compiling the
molecular data.  AMA acknowledges support from the Swedish Research Council (VR
2016-03765 and 2020-03940).  MA acknowledges funding from the Australian
Research Council through a Laureate Fellowship (FL110100012) and a Discovery
Project (DP150100250).  This research was supported by computational
resources provided by the Australian Government through the National
Computational Infrastructure (NCI) under the National Computational Merit    Allocation Scheme and the ANU Merit Allocation Scheme (project y89).  
\end{acknowledgements}

\bibliographystyle{aa} 
\bibliography{bibl.bib}
\end{document}

%% file: groups.tex
\begin{table*}
\begin{center}
\caption{Groups of indicators used in the present analysis. }
\label{tab:groups}
\begin{tabular}{c c c c c c c c}
\hline
\hline
\noalign{\smallskip}
Species & 
System & 
$|\Delta\nu|$ & 
Bands & 
$N_{\text{line}}$ & 
$\lambda_{\text{vac.}} / \nm$ & 
Param. \\ 
\hline
\hline
\noalign{\smallskip}
\ctwo & Swan & $ 0$ & (0-0), (1-1)& $   32$ & $494-516$ & \multirow{       6}{*}{$\lgeps{C}$} \\ 
\ctwo & Swan & $ 1$ & (0-1), (1-0), (1-2)& $    7$ & $473-562$ &  \\ 
\noalign{\smallskip}
CH & A-X & $ 0$ & (0-0), (1-1)& $    6$ & $425-436$ &  \\ 
CH & X-X & $ 1$ & (1-0), (2-1), (3-2)& $   48$ & $3295-3795$ &  \\ 
\noalign{\smallskip}
$\iso{12}{C}\iso{16}{O}$ & X-X & $ 1$ & (1-0), (2-1), (3-2), (4-3), (5-4)& $   28$ & $4297-6329$ &  \\ 
$\iso{12}{C}\iso{16}{O}$ & X-X & $ 2$ & (2-0), (3-1)& $   52$ & $2295-2591$ &  \\ 
\noalign{\smallskip}
\hline
\noalign{\smallskip}
NH & X-X & $ 0$ & (0-0), (1-1)& $   13$ & $11311-15023$ & \multirow{       3}{*}{$\lgeps{N}$} \\ 
NH & X-X & $ 1$ & (1-0), (2-1), (3-2), (4-3)& $   18$ & $2891-3445$ &  \\ 
\noalign{\smallskip}
CN & A-X & $ 0$ & (0-0)& $   59$ & $1088-1321$ &  \\ 
\noalign{\smallskip}
\hline
\noalign{\smallskip}
OH & X-X & $ 0$ & (0-0), (1-1), (2-2)& $   84$ & $9389-12280$ & \multirow{       3}{*}{$\lgeps{O}$} \\ 
OH & X-X & $ 1$ & (1-0), (2-1)& $   46$ & $2970-3813$ &  \\ 
OH & X-X & $ 2$ & (2-0), (3-1), (4-2)& $   15$ & $1528-1762$ &  \\ 
\noalign{\smallskip}
\hline
\hline
\end{tabular}
\end{center}
\tablefoot{Final column specifies the free parameter that is adjusted so as to match the theoretical equivalent widths to the observed ones, in the analysis that follows. \ctwo{} Swan, CH A-X and CN A-X are electronic transitions. NH X-X $|\Delta\upnu|=0$ and OH X-X $|\Delta\upnu|=0$ are rotational (or pure-rotation) transitions. The CH X-X, \comol{} X-X, NH X-X $|\Delta\upnu|=1$, OH X-X $|\Delta\upnu|\geq1$ are rovibrational (or vibration-rotation) transitions.}
\end{table*}

%% file: linelist.tex
\begin{table*}
\begin{center}
\caption{Lines used in the present analysis.}
\label{tab:linelist}
\begin{tabular}{c c c c c c c c c c c c c c}
\hline
\hline
\noalign{\smallskip}
Species & 
System & 
$|\Delta\nu|$ & 
Band & 
$\lambda_{\text{vac.}} / \nm$ & 
$E_{\text{low}} / \eV$ & 
$\lggf$ & 
$W / \mathrm{pm}$ & 
Param. & 
3D & 
\mtd{} & 
\atmo{} & 
\marcs{} & 
\handm{} \\ 
\noalign{\smallskip}
\hline
\hline
\noalign{\smallskip}
\ctwo & Swan & $ 0$ & (0-0) & $        493.812$ & $          1.112$ & $          0.843$ & $          0.530$ & $\lgeps{C}$ & $          8.401$ & $          8.415$ & $          8.380$ & $          8.376$ & $          8.459$ \\ 
\ctwo & Swan & $ 0$ & (0-0) & $        495.281$ & $          1.029$ & $          0.651$ & $          0.500$ & $\lgeps{C}$ & $          8.442$ & $          8.457$ & $          8.422$ & $          8.417$ & $          8.501$ \\ 
\ldots & \ldots & \ldots & \ldots & \ldots & \ldots & \ldots & \ldots & \ldots & \ldots & \ldots & \ldots & \ldots & \ldots \\ 
\noalign{\smallskip}
CH & A-X & $ 0$ & (1-1) & $        425.420$ & $          0.523$ & $         -1.506$ & $          3.700$ & $\lgeps{C}$ & $          8.418$ & $          8.433$ & $          8.368$ & $          8.357$ & $          8.512$ \\ 
CH & A-X & $ 0$ & (1-1) & $        425.441$ & $          0.523$ & $         -1.471$ & $          4.000$ & $\lgeps{C}$ & $          8.422$ & $          8.437$ & $          8.372$ & $          8.360$ & $          8.517$ \\ 
\ldots & \ldots & \ldots & \ldots & \ldots & \ldots & \ldots & \ldots & \ldots & \ldots & \ldots & \ldots & \ldots & \ldots \\ 
\noalign{\smallskip}
CH & X-X & $ 1$ & (1-0) & $       3294.615$ & $          0.773$ & $         -2.600$ & $          1.590$ & $\lgeps{C}$ & $          8.455$ & $          8.485$ & $          8.424$ & $          8.418$ & $          8.555$ \\ 
CH & X-X & $ 1$ & (1-0) & $       3303.095$ & $          0.915$ & $         -2.523$ & $          1.460$ & $\lgeps{C}$ & $          8.479$ & $          8.508$ & $          8.446$ & $          8.441$ & $          8.576$ \\ 
\ldots & \ldots & \ldots & \ldots & \ldots & \ldots & \ldots & \ldots & \ldots & \ldots & \ldots & \ldots & \ldots & \ldots \\ 
\noalign{\smallskip}
$\iso{12}{C}\iso{16}{O}$ & X-X & $ 1$ & (1-0) & $       4297.175$ & $          2.242$ & $         -2.876$ & $          8.240$ & $\lgeps{C}$ & $          8.482$ & $          8.580$ & $          8.550$ & $          8.540$ & $          8.623$ \\ 
$\iso{12}{C}\iso{16}{O}$ & X-X & $ 1$ & (1-0) & $       4297.866$ & $          2.286$ & $         -2.872$ & $          7.460$ & $\lgeps{C}$ & $          8.462$ & $          8.561$ & $          8.530$ & $          8.522$ & $          8.605$ \\ 
\ldots & \ldots & \ldots & \ldots & \ldots & \ldots & \ldots & \ldots & \ldots & \ldots & \ldots & \ldots & \ldots & \ldots \\ 
\noalign{\smallskip}
\hline
\noalign{\smallskip}
NH & X-X & $ 0$ & (0-0) & $      11311.232$ & $          2.138$ & $         -1.557$ & $          0.243$ & $\lgeps{N}$ & $          7.898$ & $          8.002$ & $          8.015$ & $          7.997$ & $          8.088$ \\ 
NH & X-X & $ 0$ & (0-0) & $      11471.850$ & $          2.027$ & $         -1.570$ & $          0.316$ & $\lgeps{N}$ & $          7.913$ & $          8.012$ & $          8.035$ & $          8.003$ & $          8.039$ \\ 
\ldots & \ldots & \ldots & \ldots & \ldots & \ldots & \ldots & \ldots & \ldots & \ldots & \ldots & \ldots & \ldots & \ldots \\ 
\noalign{\smallskip}
CN & A-X & $ 0$ & (0-0) & $       1087.516$ & $          0.056$ & $         -2.601$ & $          0.136$ & $\lgeps{N}$ & $          7.870$ & $          8.000$ & $          7.964$ & $          7.957$ & $          8.051$ \\ 
CN & A-X & $ 0$ & (0-0) & $       1087.548$ & $          0.089$ & $         -2.630$ & $          0.112$ & $\lgeps{N}$ & $          7.848$ & $          7.965$ & $          7.944$ & $          7.932$ & $          8.027$ \\ 
\ldots & \ldots & \ldots & \ldots & \ldots & \ldots & \ldots & \ldots & \ldots & \ldots & \ldots & \ldots & \ldots & \ldots \\ 
\noalign{\smallskip}
\hline
\noalign{\smallskip}
OH & X-X & $ 0$ & (0-0) & $       9389.310$ & $          3.070$ & $         -1.227$ & $          1.798$ & $\lgeps{O}$ & $          8.718$ & $          8.779$ & $          8.770$ & $          8.761$ & $          8.834$ \\ 
OH & X-X & $ 0$ & (0-0) & $       9391.329$ & $          3.071$ & $         -1.239$ & $          1.817$ & $\lgeps{O}$ & $          8.734$ & $          8.797$ & $          8.789$ & $          8.784$ & $          8.851$ \\ 
\ldots & \ldots & \ldots & \ldots & \ldots & \ldots & \ldots & \ldots & \ldots & \ldots & \ldots & \ldots & \ldots & \ldots \\ 
\noalign{\smallskip}
\hline
\hline
\end{tabular}
\end{center}
\tablefoot{Columns 1 to 7 show the line information, column 8 shows
the equivalent width measured in the solar disc-centre intensity,
and column 9 specifies the free parameter that is adjusted so as 
to match the theoretical equivalent widths to the observed ones.
Final columns show the elemental abundances inferred from the different 
models.  Blended transitions within the same branch were in 
some cases combined together.  
The NH lines are all in the R branch with three unresolved components.  The listed transition probabilities are for the middle component, and the measured equivalent width has been reduced by a factor of three.
Two lines of each system are shown; the full table is available online.}
\end{table*}

%% file: abundances.tex
\begin{table*}
\begin{center}
\caption{Inferred elemental abundances and uncertainties.}
\label{tab:abundances}
\begin{tabular}{c c c  c c c c c  c c c c  c}
\hline
\hline
\noalign{\smallskip}
Species & 
System & 
$|\Delta\upnu|$ & 
3D & 
\mtd{} & 
\atmo{} & 
\marcs{} & 
\handm{} & 
$\upsigma\left(\mathrm{atmos.}\right)$ & 
$\upsigma\left(\mathrm{trend}\right)$ & 
$\upsigma\left(\mathrm{comp.}\right)$ & 
$\upsigma\left(\mathrm{stat.}\right)$ & 
$\upsigma$ \\ 
\noalign{\smallskip}
\hline
\hline
\noalign{\smallskip}
\multicolumn{13}{c}{Carbon} \\
\noalign{\smallskip}
\cline{1-3}
\noalign{\smallskip}
\ctwo & Swan & 0 & $          8.455$ & $          8.473$ & $          8.438$ & $          8.433$ & $          8.520$ & $          0.018$ & $          0.003$ & $          0.006$ & $          0.005$ & $          0.020$ \\ 
\ctwo & Swan & $1$ & $          8.474$ & $          8.494$ & $          8.459$ & $          8.454$ & $          8.540$ & $          0.018$ & $          0.004$ & $          0.006$ & $          0.010$ & $          0.022$ \\ 
\noalign{\smallskip}
\cline{1-3}
\noalign{\smallskip}
\multicolumn{3}{c}{\textlangle \ctwo\textrangle} & $          8.464$ & $          8.483$ & $          8.448$ & $          8.443$ & $          8.530$ & $          0.018$ & $          0.003$ & $          0.006$ & $          0.005$ & $          0.020$ \\ 
\noalign{\smallskip}
\cline{1-3}
\noalign{\smallskip}
CH & A-X & $0$ & $          8.459$ & $          8.471$ & $          8.404$ & $          8.393$ & $          8.555$ & $          0.031$ & $          0.015$ & $          0.005$ & $          0.011$ & $          0.037$ \\ 
CH & X-X & $1$ & $          8.470$ & $          8.503$ & $          8.444$ & $          8.438$ & $          8.573$ & $          0.028$ & $          0.017$ & $          0.006$ & $          0.003$ & $          0.033$ \\ 
\noalign{\smallskip}
\cline{1-3}
\noalign{\smallskip}
\multicolumn{3}{c}{\textlangle CH\textrangle} & $          8.465$ & $          8.489$ & $          8.427$ & $          8.420$ & $          8.565$ & $          0.029$ & $          0.011$ & $          0.006$ & $          0.005$ & $          0.032$ \\ 
\noalign{\smallskip}
\cline{1-3}
\noalign{\smallskip}
\comol & X-X & $1$ & $          8.487$ & $          8.653$ & $          8.617$ & $          8.606$ & $          8.683$ & $          0.043$ & $          0.021$ & $          0.031$ & $          0.004$ & $          0.057$ \\ 
\comol & X-X & $2$ & $          8.467$ & $          8.681$ & $          8.623$ & $          8.609$ & $          8.743$ & $          0.059$ & $          0.016$ & $          0.029$ & $          0.001$ & $          0.068$ \\ 
\noalign{\smallskip}
\cline{1-3}
\noalign{\smallskip}
\multicolumn{3}{c}{\textlangle \comol\textrangle} & $          8.479$ & $          8.673$ & $          8.621$ & $          8.608$ & $          8.723$ & $          0.050$ & $          0.014$ & $          0.030$ & $          0.002$ & $          0.060$ \\ 
\noalign{\smallskip}
\cline{1-3}
\noalign{\smallskip}
\hline
\hline
\noalign{\smallskip}
\multicolumn{3}{c}{ Final} & $           8.47$ & $           8.49$ & $           8.45$ & $           8.45$ & $           8.55$ & $           0.02$ & $           0.00$ & $           0.01$ & $           0.00$ & $           0.02$ \\ 
\hline
\hline
\noalign{\smallskip}
\multicolumn{13}{c}{Nitrogen} \\
\noalign{\smallskip}
\cline{1-3}
\noalign{\smallskip}
NH & X-X & $0$ & $          7.881$ & $          8.032$ & $          8.038$ & $          8.021$ & $          8.079$ & $          0.043$ & $          0.036$ & $          0.000$ & $          0.006$ & $          0.057$ \\ 
NH & X-X & $1$ & $          7.921$ & $          7.988$ & $          7.922$ & $          7.914$ & $          8.056$ & $          0.032$ & $          0.008$ & $          0.001$ & $          0.005$ & $          0.033$ \\ 
\noalign{\smallskip}
\cline{1-3}
\noalign{\smallskip}
\multicolumn{3}{c}{\textlangle NH\textrangle} & $          7.911$ & $          8.002$ & $          7.960$ & $          7.951$ & $          8.065$ & $          0.035$ & $          0.011$ & $          0.000$ & $          0.004$ & $          0.037$ \\ 
\noalign{\smallskip}
\cline{1-3}
\noalign{\smallskip}
CN & A-X & $0$ & $          7.867$ & $          7.964$ & $          7.933$ & $          7.927$ & $          8.016$ & $          0.031$ & $          0.008$ & $          0.030$ & $          0.003$ & $          0.044$ \\ 
\noalign{\smallskip}
\cline{1-3}
\noalign{\smallskip}
\multicolumn{3}{c}{\textlangle CN\textrangle} & $          7.867$ & $          7.964$ & $          7.933$ & $          7.927$ & $          8.016$ & $          0.031$ & $          0.008$ & $          0.030$ & $          0.003$ & $          0.044$ \\ 
\noalign{\smallskip}
\cline{1-3}
\noalign{\smallskip}
\hline
\hline
\noalign{\smallskip}
\multicolumn{3}{c}{ Final} & $           7.89$ & $           7.99$ & $           7.95$ & $           7.94$ & $           8.04$ & $           0.03$ & $           0.01$ & $           0.01$ & $           0.00$ & $           0.04$ \\ 
\hline
\hline
\noalign{\smallskip}
\multicolumn{13}{c}{Oxygen} \\
\noalign{\smallskip}
\cline{1-3}
\noalign{\smallskip}
OH & X-X & $0$ & $          8.690$ & $          8.792$ & $          8.795$ & $          8.782$ & $          8.837$ & $          0.031$ & $          0.041$ & $          0.005$ & $          0.001$ & $          0.052$ \\ 
OH & X-X & $1$ & $          8.707$ & $          8.800$ & $          8.737$ & $          8.728$ & $          8.884$ & $          0.039$ & $          0.008$ & $          0.004$ & $          0.002$ & $          0.040$ \\ 
OH & X-X & $2$ & $          8.690$ & $          8.756$ & $          8.685$ & $          8.673$ & $          8.848$ & $          0.038$ & $          0.066$ & $          0.003$ & $          0.012$ & $          0.077$ \\ 
\noalign{\smallskip}
\cline{1-3}
\noalign{\smallskip}
\multicolumn{3}{c}{\textlangle OH\textrangle} & $          8.699$ & $          8.791$ & $          8.756$ & $          8.751$ & $          8.852$ & $          0.036$ & $          0.017$ & $          0.004$ & $          0.002$ & $          0.040$ \\ 
\noalign{\smallskip}
\cline{1-3}
\noalign{\smallskip}
\hline
\hline
\noalign{\smallskip}
\multicolumn{3}{c}{ Final} & $           8.70$ & $           8.79$ & $           8.76$ & $           8.75$ & $           8.85$ & $           0.04$ & $           0.02$ & $           0.00$ & $           0.00$ & $           0.04$ \\ 
\hline
\hline
\end{tabular}
\end{center}
\tablefoot{Abundances are shown for different groups and species, 
for all of the models considered in the present analysis.
Also shown is the breakdown of uncertainties for the 3D model, with the overall
uncertainty given in the final column.}
\end{table*}